\documentclass[preprint]{aastex}
\usepackage{graphicx}
\usepackage{epstopdf}

\shorttitle{M31 globulars and X--ray binaries}
\shortauthors{Agar \& Barmby}

\begin{document}

\defcitealias{peacock10}{P10}
\defcitealias{king66}{K66}

\title{M31 globular cluster structures and the presence of X--ray binaries%
\footnote{
Based on observations made with the NASA/ESA Hubble Space Telescope, and 
obtained from the Hubble Legacy Archive, which is a collaboration between the 
Space Telescope Science Institute (STScI/NASA), the Space Telescope European 
Coordinating Facility (ST-ECF/ESA) and the Canadian Astronomy Data Centre (CADC/NRC/CSA). }
}

\author{J.R.R. Agar \& P. Barmby\altaffilmark{1}}
\affil{Department of Physics \& Astronomy, University of Western Ontario, London, ON N6A 3K7, Canada}
\altaffiltext{1}{Corresponding author, pbarmby@uwo.ca}

\begin{abstract}
The Andromeda galaxy, M31, has several times the number of globular clusters found in the Milky Way.
It contains a correspondingly larger number of low mass X--ray binaries (LMXBs) associated with globular clusters,
and as such can be used to investigate the cluster properties which lead to X--ray binary formation.
The best tracer of the spatial structure of M31 globulars is the high-resolution imaging available
from the \textit{Hubble Space Telescope (HST)}, and we have used \textit{HST} data to derive structural parameters for 29 
LMXB-hosting M31 globular clusters. These measurements are combined  with structural parameters
from the literature for a total of 41 (of 50 known) LMXB clusters and a comparison sample of 65 non-LMXB 
clusters. Structural parameters measured in blue bandpasses are found to be slightly
different (smaller core radii and higher concentrations) than those measured in red bandpasses;
this difference is enhanced in LMXB clusters and could be related to stellar population differences.
Clusters with LMXBs show higher collision rates for their mass compared to clusters without LMXBs
and collision rates estimated at the core radius show larger offsets than rates estimated at the half-light radius.
These results are consistent with the dynamical formation scenario for LMXBs.
A logistic regression analysis finds that, as expected, the probability of a cluster hosting an LMXB
increases with increasing collision rate and proximity to the galaxy center. The same analysis
finds that probability of a cluster hosting an LMXB  decreases with increasing
cluster mass at a fixed collision rate, although we caution that this could be due to sample selection effects.
Metallicity is found to be a less important predictor of LMXB probability
than collision rate, mass, or distance, even though LMXB clusters  have a higher metallicity on average.
This may be due to the interaction of location and metallicity: a sample of M31 LMXBs with a greater
range in galactocentric distance would likely contain more metal-poor clusters and make it possible
to disentangle the two effects.

\end{abstract}


\section{Introduction}

The high stellar density and age of globular clusters (GCs)  make them among
the most likely places in the Universe to find the products of stellar dynamical interactions.
Among such products are Low Mass X--ray Binary systems (LMXBs), binary systems where one of the components is a compact 
object, such as a neutron star or black hole. There are several theories as to how these systems form, including capture 
by remnants of massive stars \citep{clark75}, tidal capture by neutron stars in close encounters by main sequence stars 
\citep{fabian75} or direct collisions between giants and neutron stars \citep{sutantyo75}. 
Early work on X--ray sources in the Milky Way found that $\sim10$\% of luminous X--ray sources were located in GCs. 
This observation implies that the probability per unit mass of finding an LMXB is 200--300 times higher in GCs 
than in the rest of the Galaxy \citep{verbunt06}. The high 
occurrence of LMXBs along with the high central density of GCs leads to the hypothesis that
LMXBs  should be located close to the center of the cluster. This was confirmed by \citet{jernigan79}, who 
measured the positions of the LMXBs in GCs and found that the positions of the LMXBs correspond to 
the center of the clusters within 30 arcsec \citep{verbunt06}.

The small number of Galactic globulars with LMXBs, combined with obscuration
by dust in the Milky Way's disk, means that it is difficult to correlate the properties of LMXBs
with those of their host clusters. The next logical step is to look at nearby galaxies.
However, even in nearby galaxies  instruments with very good spatial resolution 
are needed in order to associate  LMXBs with  GCs. The advent of the \textit{Chandra} 
X--ray Observatory  made it possible to study  X--ray sources  
in many nearby galaxies out to distances of 20--30 Mpc  \citep{fabbiano06}.

The nearest large galaxy, M31 or the Andromeda Galaxy, has had both its X--ray source and
GC populations extensively characterized. Andromeda  contains over 2000 GC 
candidates \citep{galleti09} with over 400 of these now confirmed as GCs \citep{caldwell11}.
Among the first attempts to identify X--ray sources in M31 was the ROSAT PSPC survey of
\citet{supper01}. This survey was not limited to 
GC sources, but instead studied the brightest X--ray sources in the galaxy. 
The survey identified 560 X--ray sources in a 10.7~deg$^2$ 
field of view with fluxes  $7 \times 10^{-15}$~erg~cm$^{-2}$~s$^{-1}$ to $7.6 \times 10^{-12}$~erg~cm$^{-2}$~s$^{-1}$
(corresponding to luminosities of $5\times10^{35}-5.5\times10^{38}$ erg~s$^{-1}$). 
Of these, 33 were identified as with GCs. The 10 brightest sources were identified as 
belonging to GCs, and all but one showed the spectrum of an LMXB system 
\citep{trinchieri99}.

Both \textit{XMM-Newton}  and \textit{Chandra} observations were used in the next generation of
studies of LMXBs in M31 GCs. \citet{distefano02} used \textit{Chandra} observations to 
survey 2560 square arcmin  and found that the brightest sources in the majority of their  observations resided in GCs. 
28 GC LMXBs were studied, 15 of which were newly 
discovered. The authors identified two main differences between the M31 and Galactic populations: the 
peak X--ray luminosity of sources in M31 is higher, and the high end of the 
X--ray luminosity distribution function is more populated in M31 than in the Milky Way. 
Using both \textit{XMM-Newton}  and \textit{Chandra}  data, \citet{tp04} surveyed approximately   
6100 square arcminutes of M31 and found 43 LMXBs which were associated with a GC. The 
brightest sources tended to reside at large galactocentric radii and showed spectral properties of LMXBs, 
consistent with the findings of \citet{trinchieri99}.  GCs hosting bright LMXBs  tended 
to be optically brighter and more metal-rich than non-hosting GCs; however, the
brightest sources ($L_X>10^{38}$~erg~s$^{-1}$) tended to reside in more metal-poor clusters.

Taking advantage of more definitive catalogs of M31 GCs in the optical and near-infrared,  and
more complete coverage with \textit{XMM-Newton}, \citet[hereafter P10]{peacock10} updated the comparison of
LMXB- and non-LMXB-containing M31 clusters. 
They identified 41 LMXBs with confirmed old clusters, 3 of these being newly identified, and 
showed that LMXBs preferred brighter (i.e., more massive) clusters, as well as those with higher stellar collision rates $\Gamma$.%
\footnote{$\Gamma$ is also called the `encounter rate' by some authors.} 
\citetalias{peacock10} also 
 found that clusters in which LMXBs resided showed a higher than average stellar collision rate for their mass
and suggested that metallicity  effects could not explain this phenomenon. 
They concluded that a high stellar collision rate is the primary indicator 
of the likelihood of a LMXB being present in any particular M31 GC,
consistent with the   dynamical formation scenario for LMXBs.
Most recently,  variability analysis of \textit{Chandra} observations of 34 M31 GCs and candidates with X--ray sources led
\citet{barnard12} to conclude that all of the X--ray sources were likely LMXBs.

The question of whether  $\Gamma$  is linearly proportional to the LMXB-hosting probability has 
implications for LMXB formation and destruction channels.  As discussed by  \citet{maccarone11}, a shallower-than-linear
relation can imply that LMXBs are destroyed in the densest clusters, and some previous work has found
such a shallower dependence \citep{jordan04a,sivakoff07,jordan07}. However, \citet{maccarone11}  showed that
computations of $\Gamma$ are sensitive to the exact structural parameters used.
They found that computing  the collision rate using cluster parameters measured at the half-light radius $r_h$ rather than the core,
or using core parameters with significant measurement errors, can both result in shallower-than-linear relations.
Comparing the collision rate proxy $\Gamma=\rho_0^{3/2} r_c^2$  to the half-light version $\Gamma_h=\rho_h^{3/2} r_h^2$ for M31 clusters,
\citet{maccarone11} found the former to be ``a much better predictor of whether a cluster will host 
an X--ray source.''

As \citet{maccarone11} pointed out, M31 has the only extragalactic GCs for which typical cluster cores are resolved
by \textit{Hubble Space Telescope} (\textit{HST}) imaging. This galaxy therefore provides an important bridge between
the studies of LMXBs in Galactic globulars and more distant ellipticals. 
Here we build on two recent studies of M31 GCs: \citetalias{peacock10}, who derived a comprehensive list of LMXB-containing GCs
in M31 and studied their structural parameters in ground-based near-infrared imaging  \citep{peacock09}, 
and \citet{barmby07}, who used \textit{HST}-derived surface brightness
profiles to show that M31 GCs fall on a fundamental plane similar to those in other galaxies.
With a larger sample of clusters than in \citet{barmby07}, and better spatial resolution than
\citet{peacock09} had available, we can check the comparison between M31 LMXB- and non-LMXB clusters
as well as  between collision rate proxies. The large, well-characterized M31 GC system is an excellent
cross-check  for conclusions derived from the study  of the LMXB-GC connection in more distant, and often less typical, galaxies.

\section{Observational data and reduction}

\subsection{Cluster sample and archival data}
\label{sec:sample}

As the starting point for our analysis,
we use the matched list of M31 GCs hosting LMXBs compiled by \citetalias{peacock10}.
This work  combined recent studies on M31 GC LMXBs
using \textit{ROSAT} \citep{supper01},  \textit{Chandra} \citep{tp04}
and \textit {XMM-Newton} \citepalias{peacock10}. We include all 45 sources listed in Table 3 of \citetalias{peacock10},
all of which are classified as old clusters, and the 10 objects listed as candidate clusters from Table 4
(but see below).
Of this second group, two objects (B138 and NB16) were later classified by \citet{caldwell11} as old clusters.
To the list of 55 objects from \citetalias{peacock10} we add NB21, which was listed
as a GC X--ray source by \citet{supper01} and is a confirmed old globular, 
but is not mentioned in any of the later papers.
No X--ray luminosity cuts were applied to the sample. The 2XMMi catalog which \citetalias{peacock10}
matched against globular cluster positions has a homogeneous detection limit of $L_X\sim 10^{36}$~erg~s$^{-1}$,
but a few sources are fainter than this limit.
Of the final list of potential 56 GC LMXBs, 12 have published \textit{HST} structural analyses
from \citet{barmby07} or \citet{barmby02}.
We searched the Hubble Legacy Archive (HLA) in June 2012 for WFPC2 or ACS
images  of the remaining clusters and found available data for 35 additional objects. 
About half of these are from the Panchromatic Hubble Andromeda Treasury \citep[PHAT;][]{dalcanton12}
while the remainder are from other programs. None of these programs specifically targeted the
GCs. M31 is large compared to the \textit{HST} instruments' fields of view, and
no image contains more than one LMXB-hosting GC.

\begin{figure}
\plotone{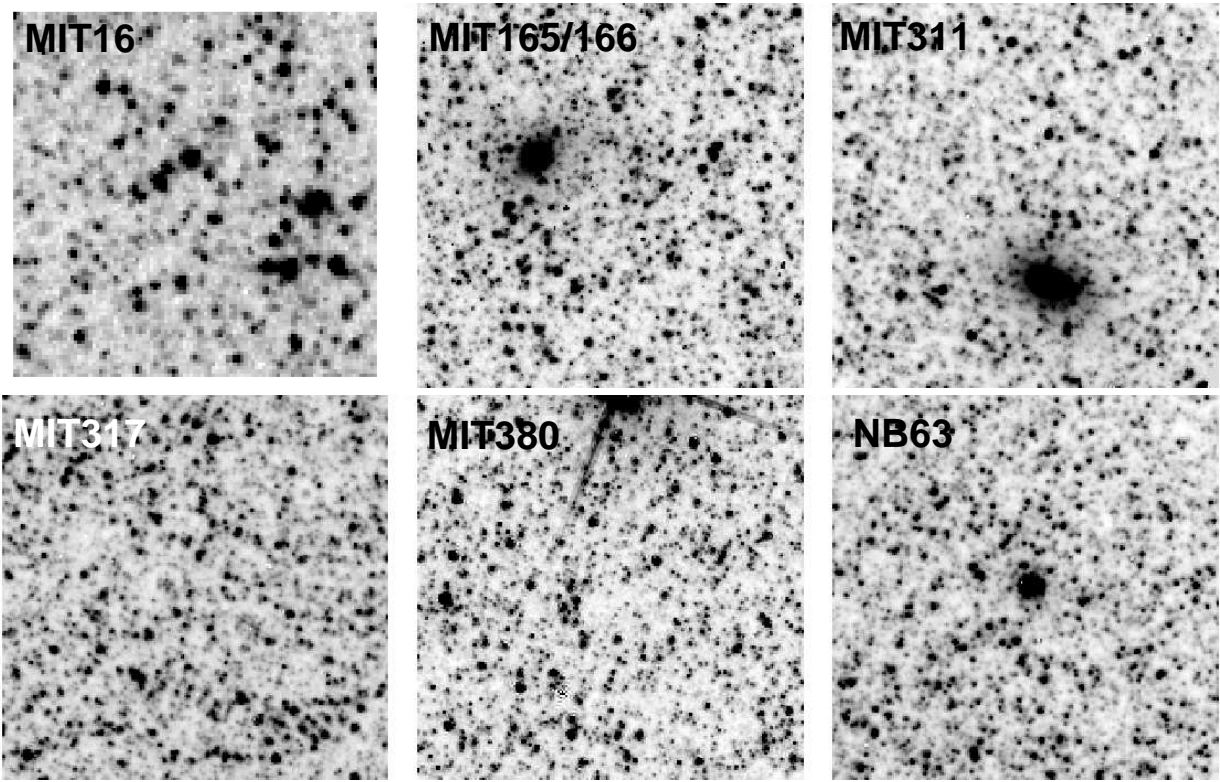}
\caption{\textit{HST} images of LMXB-host cluster candidates which are not 
confirmed clusters.
Each image is 10 arcseconds across, centered on the cluster position
as given by \citet{peacock10} with north up and east left.
All images are in the F814W filter with the MIT16 image taken by WFPC2
and all others by ACS.
\label{fig:noncl}}
\end{figure}

A total of six objects listed in Table~4 of \citetalias{peacock10} as `possible clusters' have \textit{HST} data but
do not appear to be true clusters. \textit{HST} images of all six are shown in Figure~\ref{fig:noncl}. MIT16 is in a very 
crowded field with no obvious cluster-like object nearby. There are extended objects within a 
few arcseconds of the coordinates of MIT165/166 and MIT311, but neither appears to be
resolved into individual stars.
Neither of these has counterparts in other published catalogs of M31 star clusters.
There are no obvious cluster-like objects near the coordinates of MIT317 and MIT380.
NB63 appears to be a star, consistent with its entry in the online catalog of Caldwell.
This leaves a total of 50 LMXB-hosting GCs or candidates, of which 48 are confirmed clusters
and 41 have \textit{HST} imaging. The clusters newly analyzed here, and the properties
of the relevant \textit{HST} images, are listed in Table~\ref{tab:m31hst}. Cross-references for cluster names and
positions can be found in the Revised Bologna Catalog of M31 Globular Clusters \citep[RBC;][]{rbc07}.

\begin{deluxetable}{lllllrrl}
\tabletypesize{\small}
\tablewidth{0pt}
\tablecaption{New \textit{HST} data for LMXB globular clusters \label{tab:m31hst}}
\tablehead{
\colhead{Cluster} & \colhead{Camera} &  \colhead{Filter} & \colhead{Exposure time\tablenotemark{a}} & \colhead{$E(B-V)$\tablenotemark{b}} 
& \colhead{$(V-m)_0$\tablenotemark{c}}
& \colhead{[Fe/H]\tablenotemark{d}} &  \colhead{$M/L_V$\tablenotemark{e}} \\
\colhead{} & \colhead{} & \colhead{} &\colhead{s} & \colhead{mag} & \colhead{mag} & \colhead{dex} & \colhead{M$_{\sun}$/L$_{\sun}$}
}

\startdata
\object[SKHV 148]{B086-G148}                &   ACS & F435W &  \dataset[j9ud17rlq]{12210}& 0.15 &$-0.67$& $-1.82$& 1.88\\
\object[Bol D091]{B091D-D057}               &   ACS & F555W &  \dataset[j92ga7vlq]{151}  & 0.24 &$-0.05$& $-0.70$& 2.44\\
\object[Bol D091]{B091D-D057}               &   ACS & F814W &  \dataset[j92ga7vmq]{457}  & 0.24 &$ 1.00$& $-0.70$& 2.44\\
\object[2MASXi J0042250+405717]{B094-G156}  &   ACS & F555W &  \dataset[j92gb9brq]{413}  & 0.07 &$-0.06$& $-0.40$& 2.90\\
\object[NBol 002]{B096-G158}                &   ACS & F814W &  \dataset[jbf301hmq]{3200} & 0.26 &$ 1.14$& $-0.28$& 3.12\\
\object[NBol 001]{B107-G169}                &   ACS & F475W &  \dataset[jbf301i0q]{3600} & 0.28 &$-0.30$& $-0.97$& 2.16\\
\object[NBol 001]{B107-G169}                &   ACS & F814W &  \dataset[jbf301hmq]{3200} & 0.28 &$ 0.91$& $-0.97$& 2.16\\
\object[SKHV 178]{B110-G172}                & WFPC2 & F606W &  \dataset[u581r201r]{1800} & 0.20 &$ 0.27$& $-0.66$& 2.49\\
\object[SKHV 178]{B116-G178}                &   ACS & F555W &  \dataset[j92ga3dkq]{222}  & 0.62 &$-0.06$& $-0.64$& 2.52\\
\object[SKHV 178]{B116-G178}                &   ACS & F814W &  \dataset[j92ga3dlq]{457}  & 0.62 &$ 1.05$& $-0.64$& 2.52\\
\object[SKHV 176]{B117-G176}                & WFPC2 & F336W &  \dataset[u92g1202m]{4000} & 0.04 &$-1.06$& $-1.72$& 1.88\\
\object[SKHV 187]{B128-G187}                & WFPC2 & F555W &  \dataset[u92g1202m]{160}  & 0.15 &$-0.03$& $-0.56$& 2.63\\
\object[SKHV 187]{B128-G187}                & WFPC2 & F814W &  \dataset[u92g1201m]{160}  & 0.15 &$ 0.89$& $-0.56$& 2.63\\
\object[SKHV 192]{B135-G192}                &   ACS & F555W &  \dataset[j92ga7vmq]{151}  & 0.27 &$-0.05$& $-1.82$& 1.88\\
\object[SKHV 192]{B135-G192}                &   ACS & F814W &  \dataset[j92ga7vmq]{457}  & 0.27 &$ 0.87$& $-1.82$& 1.88\\
\object[C401141021]{B138}                   &   ACS & F475W &  \dataset[jbf301i0q]{1890} & 0.22 &$-0.46$& $-0.04$& 3.72\\
\object[C401540596]{B144}                   &   ACS & F475W &  \dataset[jbf301i0q]{3600} & 0.05 &$-0.53$& $ 0.08$& 3.96\\
\object[C401540596]{B144}                   &   ACS & F814W &  \dataset[jbf301hmq]{3200} & 0.05 &$ 1.26$& $ 0.08$& 3.96\\
\object[C401840589]{B146}                   &   ACS & F475W &  \dataset[jbf301i0q]{3600} & 0.06 &$-0.49$& $-1.01$& 2.13\\
\object[C401840589]{B146}                   &   ACS & F814W &  \dataset[jbf301hmq]{3200} & 0.06 &$ 1.01$& $-1.01$& 2.13\\
\object[SKHV 200]{B148-G200}                &   ACS & F475W &  \dataset[jbf301i0q]{3600} & 0.17 &$-0.38$& $-1.09$& 2.07\\
\object[SKHV 200]{B148-G200}                &   ACS & F814W &  \dataset[jbf301hmq]{3200} & 0.17 &$ 0.80$& $-1.09$& 2.07\\
\object[SKHV 203]{B150-G203}                & WFPC2 & F555W &  \dataset[ua250301m]{320}  & 0.28 &$-0.03$& $-0.28$& 3.12\\
\object[SKHV 203]{B150-G203}                & WFPC2 & F814W &  \dataset[ua250303m]{400}  & 0.28 &$ 0.88$& $-0.28$& 3.12\\
\object[Bol 153]{B153}                      &   ACS & F475W &  \dataset[jbf301i0q]{3600} & 0.05 &$-0.55$& $-0.28$& 3.12\\
\object[Bol 153]{B153}                      &   ACS & F814W &  \dataset[jbf301i0q]{3200} & 0.05 &$ 1.23$& $-0.28$& 3.12\\
\object[Bol 159]{B159}                      &   ACS & F435W &  \dataset[j8vp08jmq]{2000} & 0.36 &$-0.76$& $-1.17$& 2.02\\
\object[Bol 159]{B159}                      &   ACS & F475W &  \dataset[jbfb13020]{1900} & 0.36 &$-0.41$& $-1.17$& 2.02\\
\object[Bol 159]{B159}                      &   ACS & F814W &  \dataset[jbfb13020]{1715} & 0.36 &$ 0.94$& $-1.17$& 2.02\\
\object[SKHV 215]{B161-G215}                &   ACS & F475W &  \dataset[jbf301i0q]{3600} & 0.17 &$-0.38$& $-1.06$& 2.09\\
\object[SKHV 215]{B161-G215}                &   ACS & F814W &  \dataset[jbf301hmq]{3200} & 0.17 &$ 0.88$& $-1.06$& 2.09\\
\object[SKHV 217]{B163-G217}                & WFPC2 & F555W &  \dataset[u92g0602m]{40}   & 0.14 &$-0.03$& $-0.13$& 3.48\\
\object[SKHV 217]{B163-G217}                & WFPC2 & F814W &  \dataset[u92g0601m]{40}   & 0.14 &$ 1.14$& $-0.13$& 3.48\\
\object[Bol 164]{B164-V253}                 &   ACS & F475W &  \dataset[jbf306o6q]{3600} & 0.12 &$-0.54$& $-0.29$& 3.10\\
\object[Bol 164]{B164-V253}                 &   ACS & F814W &  \dataset[jbf306nrq]{3200} & 0.12 &$ 1.15$& $-0.29$& 3.10\\
\object[SKHV 233]{B182-G233}                &   ACS & F555W &  \dataset[j92gc6d0q]{141}  & 0.25 &$-0.05$& $-1.03$& 2.11\\
\object[SKHV 233]{B182-G233}                &   ACS & F814W &  \dataset[j92gc6d1q]{457}  & 0.25 &$ 1.00$& $-1.03$& 2.11\\
\object[SKHV 235]{B185-G235}                &   ACS & F435W &  \dataset[j96q03ifq]{4476} & 0.11 &$-0.90$& $-0.61$& 2.56\\
\object[2MASX J00434551+4136578]{B193-G244} &   ACS & F475W &  \dataset[jbf101uzq]{3600} & 0.11 &$-0.50$& $-0.11$& 3.56\\
\object[SKHV 254]{B204-G254}                &   ACS & F475W &  \dataset[jbfb16i7q]{3600} & 0.12 &$-0.47$& $-0.69$& 2.45\\
\object[SKHV 254]{B204-G254}                &   ACS & F814W &  \dataset[jbfb16hsq]{3200} & 0.12 &$ 1.01$& $-0.69$& 2.45\\
\object[SKHV 264]{B213-G264}                &   ACS & F435W &  \dataset[j96q05swq]{2925} & 0.15 &$-0.98$& $-0.77$& 2.35\\
\object[SKHV 011]{B293-G011}                & WFPC2 & F555W &  \dataset[u4ca5501r]{5300} & 0.04 &$-0.03$& $-1.80$& 1.88\\
\object[SKHV 011]{B293-G011}                & WFPC2 & F606W &  \dataset[u9x6f001m]{12100}& 0.04 &$ 0.22$& $-1.80$& 1.88\\
\object[SKHV 011]{B293-G011}                & WFPC2 & F814W &  \dataset[u9x6f002m]{12100}& 0.04 &$ 0.81$& $-1.80$& 1.88\\
\object[SKHV 307]{B375-G307}                &   ACS & F475W &  \dataset[jbf806leq]{3600} & 0.29 &$-0.33$& $-0.90$& 2.22\\
\object[SKHV 307]{B375-G307}                &   ACS & F814W &  \dataset[jbf806kfq]{3200} & 0.29 &$ 0.81$& $-0.90$& 2.22\\
\object[CXO J004246.0+411736]{BH16}         &   ACS & F435W &  \dataset[j8vp08jmq]{2200} & 0.13 &$-0.80$& $-1.00$& 2.13\\
\object[NBol 021]{NB21}                     &   ACS & F435W &  \dataset[j8vp09aiq]{2200} & 0.02 &$-0.51$& $-1.13$& 2.05\\
\object[NBol 021]{NB21}                     &   ACS & F475W &  \dataset[jbf308dzq]{1890} & 0.02 &$-0.28$& $-1.13$& 2.05\\
\object[NBol 021]{NB21}                     &   ACS & F814W &  \dataset[jbf308diq]{1700} & 0.02 &$ 1.07$& $-1.13$& 2.05\\
\enddata
\tablenotetext{a}{The electronic edition contains links to the original {\em HST\/} dataset.}
\tablenotetext{b}{Reddening values are from \citet{fan10}, except for B091D,BH16,NB21 from \citet{caldwell11}, B159 from \citet{fan08}; see text.}
\tablenotetext{c}{Extinction-corrected color used to convert measurements to the $V$ band; see text.}
\tablenotetext{d}{Metallicities are from \citet{caldwell11}.}
\tablenotetext{e}{Computed using metallicity-dependent $M/L_V$ for an age of 13~Gyr, see text.}
\end{deluxetable}

\clearpage

\clearpage

\begin{figure}
\plotone{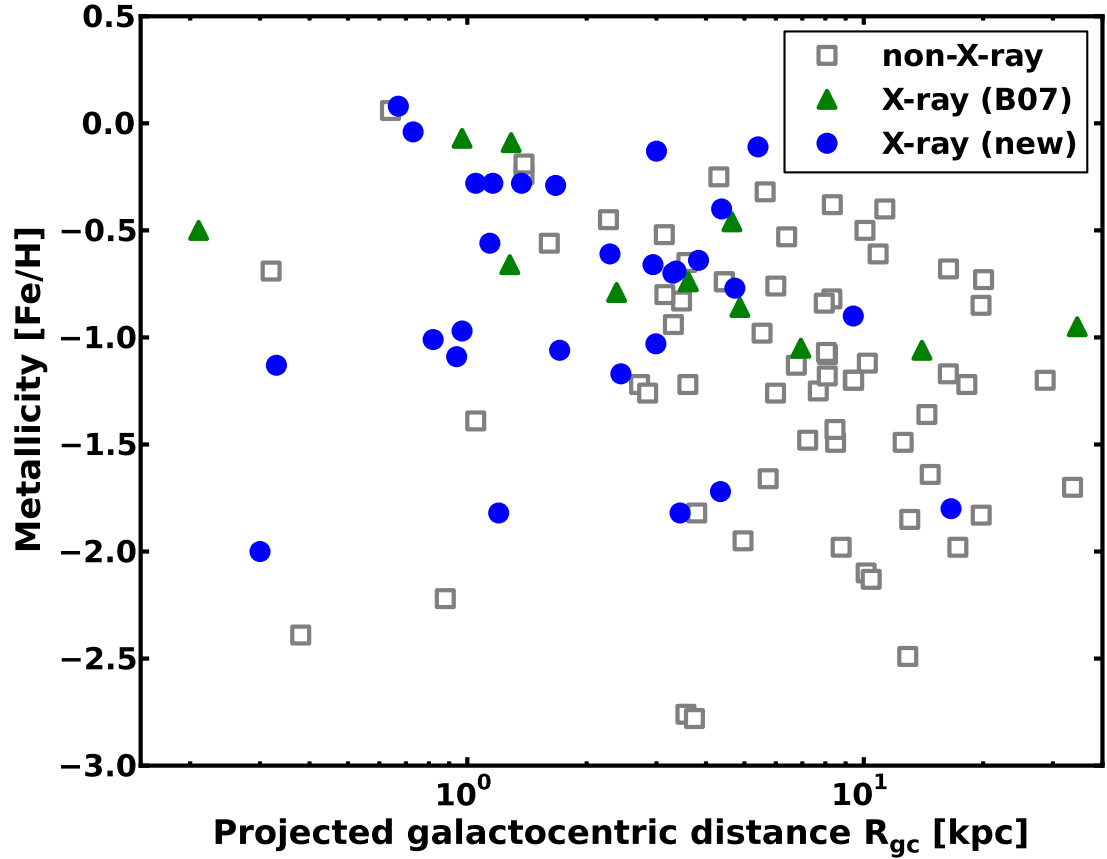}
\caption{Comparison of projected galactocentric distance and metallicity for
LMXB- and non-LMXB-hosting M31 clusters. The LMXB clusters
have higher metallicities and smaller projected distances, due in part
to the large number of LMXB clusters newly imaged by the PHAT project.
\label{fig:x_compare}}
\end{figure}

Our total sample of M31 GCs with both LMXBs and \textit{HST} imaging
comprises 41 objects.
This is about twice as large as the number of clusters with 
LMXBs and structural parameters analyzed by \citet{peacock09,peacock10};
the overlap between the two samples is 15 objects, or about 40\% of our sample.
(The near-IR imaging used by \citealt{peacock09} covered only the disk of M31 and therefore only about half of the LMXB-containing clusters).
The comparison sample for the LMXB clusters includes the 65 old, non-LMXB-hosting 
clusters from  \citet{barmby07}, which includes a re-analysis of the clusters studied in  \citet{barmby02}, and \citet{barmby09a}.%
\footnote{After this paper was submitted, \citet{wang13} re-analyzed surface brightness profiles
for most of these clusters. Their sample  includes only a few of the clusters listed in Table~\ref{tab:m31hst},
so to avoid introducing any systematic differences between the X--ray and non-X--ray samples, we
do not use their results.}
Most of the comparison sample is from \citet{barmby07}, so we abbreviate the reference to the previous studies as B07.
Both the LMXB and non-LMXB samples contain both objects specifically targeted for  \textit{HST} observations 
and clusters serendipitously observed as part of other programs. 
Unlike \citetalias{peacock10}, these samples do not cover
a contiguous region of the galaxy, nor are they complete to some limiting magnitude. 
Therefore an important question is whether our conclusions are likely to be affected by selection bias.
At least some of the programs which targeted M31 GCs with \textit{HST} 
\citep[e.g.][]{ajhar96,meylan01,brown04,barmby07,perina11,tanvir12}
favored massive clusters but, to our knowledge, none of them used the presence of
an LMXB as a selection criterion.%
\footnote{The long-term \textit{HST} program carried out by M. Garcia and collaborators to follow-up M31 X--ray novae 
targets field stars and not clusters (R. Barnard, personal communication).}
In this work we did not search the \textit{HST} archive for additional non-LMXB clusters,
and the LMXB and non-LMXB cluster samples do differ in both galactocentric position and
metallicity (see Figure~\ref{fig:x_compare}). 
Since galactocentric position and metallicity are correlated, and both
 are known to have small but systematic effects on cluster radii \citep{barmby02,barmby07},
this is an important complication to our analysis and will be further discussed in Section~\ref{sec:disc}.

For the clusters with newly available \textit{HST} data, additional cluster properties are needed to 
convert the observed flux-based measurements to luminosities
and mass-linked quantities.  To convert fluxes to luminosities, we assume a distance to M31 of 783~kpc \citep{stanek98}, 
for which 1\arcsec\ corresponds to 3.797~pc.
To correct for extinction, we use the extinction coefficients given by \citet{girardi08}
and the values of $E(B-V)$ given by \citet{fan10}, with a few exceptions. \citet{fan10} find
cluster NB21 to be young and heavily-reddened, while \citet{caldwell11} find it
to be old with little extinction. For this object, we adopt the \citet{caldwell11} 
extinction. Clusters B159 and BH16 do not have reddening values given by \citet{fan10}; we adopt the
B159 value from \citet{fan08} and the BH16 value from \citet{caldwell11}.

Our sample of clusters was observed in various filters but we convert
all of these to the $V$-band for inter-comparison.
This was done using the same method used as in \citet{barmby07,barmby09a}:
\textit{HST} transformations \citep{sirianni05,holtzman95} and ground-based integrated colors.
For example, equation 12 and Table 22 of \citet{sirianni05} give the transformation from magnitudes in the F475W
band to $B$ as:
\begin{equation}
B=[26.146-2.5\log(e^{-}\,\,{\rm  s^{-1}})]+0.389(B-V)+0.032(B-V)^2
\end{equation}
where all quantities are extinction corrected and the quantity in square brackets is the Vega magnitude in F475W.
This can be re-arranged to give 
\begin{equation}
V-m_{\rm F475W} = 26.146-0.611(B-V) +0.032(B-V)^2
\end{equation}
We use ground-based colors from Version 5 of the RBC, except in a few
cases (B148, B375, NB21) where the RBC colors seemed to be far too blue for the spectroscopic metallicity.
For B148 and B375 we adopted the \citet{fan10} colors; for NB21 we measured an F435W magnitude
from the HST image and used this to compute $B-V$.
The extinction-corrected colors $(V-x)_0$, where $x$ is the observed-band magnitude,
are tabulated in Table~\ref{tab:m31hst}. Uncertainties of 0.1~mag in $(V-x)_0$ are assumed
and propagated through the parameter estimates.

To convert  $V$-band luminosity to mass, two options are available:
using the predictions of  population synthesis models, or directly measured
dynamical mass-to-light ratios. The previous work from which our parent
sample of M31 clusters was drawn \citep{barmby07} used the first method, applying 
metallicity-dependent mass-to-light ratios from the population synthesis models of \citet{bruzual03} with
a \citet{chabrier03} IMF. \citet{mclaughlin05} and \citet{mclaughlin08a} used the same models.
Recently \citet{strader11} have measured spectroscopic  mass-to-light ratios for 163 M31 GCs,
including 20 of the LMXB-hosting clusters considered here.
Because only about half of our sample have direct measurements of $M/L$,
we chose to retain the population synthesis model approach. Although there are indications that model-predicted values may
not be correct \citep[see][]{strader11}, this approach has the advantage of
being internally consistent for all of the sample. 
As a source of metallicity values, we use  \citet{caldwell11}, which gives
values for all clusters except BH16. For this cluster, we assume a typical
value ${\rm [Fe/H]}=-1.0$. Metallicities and the resulting $M/L_V$ are tabulated 
in Table~\ref{tab:m31hst}. As for color transformations, assumed uncertainties of 10\% in $M/L_V$ 
are propagated through the resulting  parameter estimates.

\subsection{Surface Brightness Profiles}

The images analyzed for this project were taken directly from the HLA. The PHAT data processing
is described by \citet{dalcanton12}; the non-PHAT data have the standard HLA 
processing (combining for cosmic-ray removal, drizzling) applied.
Where clusters were observed in more than one filter, we analyzed images in
the two filters closest to the $V$-band, as a check on our results.
A number of clusters had images in more than one filter, but with the image in the redder
filter showing detector saturation in the cluster core; these saturated images were not analyzed.
Image analysis used the same method described in \citet{barmby07} and we refer the reader
to that paper for the details. Briefly, surface brightness profiles were measured on
circular annuli using  the {\sc ellipse} task in IRAF, converted from image counts to solar luminosities
per square parsec (L$_{\sun}$~pc$^{-2}$), and fit to models using the {\sc gridfit} code
described by \citet{mclaughlin05}.

The conversion between image counts and solar luminosities per square parsec
depends on the image zeropoint, instrument pixel scale, and the solar absolute magnitude
in the specific bandpass.  
The calculation of the conversion factor is described in  \citet{barmby09a} 
and the values used here are tabulated in  Table~\ref{tab:zps} for reference.
While the solar absolute magnitude for most filters was taken from the tabulation by C. Willmer%
\footnote{\url{http://mips.as.arizona.edu/$\sim$cnaw/sun.html}},
for the F336W filter we computed this parameter using  SYNPHOT in IRAF,  
deriving  M$_{\sun, {\rm F336W}}=5.58$.

\begin{deluxetable}{llll}
\tablewidth{0pt}
\tablecaption{Zeropoints and Conversion Factors\label{tab:zps}}
\tablehead{
\colhead{Camera} & \colhead{Filter} & \colhead{Zeropoint\tablenotemark{a}} & \colhead{Conversion Factor\tablenotemark{b}}}
\startdata
ACS & F435W & 25.676 & 1433.0455\\
ACS & F475W & 26.154&719.548 \\
ACS & F555W & 25.74383 & 746.66878 \\
ACS & F814W & 25.53561 & 474.69646\\
WFPC2/WFC & F555W & 24.676& 499.11429\\
WFPC2/WFC &F606W &25.017 &306.05537 \\
WFPC2/WFC & F814W &23.774 &601.17374 \\
WFPC2/PC & F336W & 21.548&69736.475 \\
WFPC2/PC & F555W &24.664 & 2018.6452\\
WFPC2/PC & F606W & 25.006&1236.6876 \\
WFPC2/PC & F814W & 23.758&2440.3942 \\
\enddata
\tablenotetext{a}{Zeropoint for conversion from counts~s$^{-1}$~arcsec$^{-2}$ to Vega mag~arcsec$^{-2}$} 
\tablenotetext{b}{Multiplicative surface brightness conversion factor from counts~s$^{-1}$~pixel$^{-1}$ to solar luminosities per square parsec.}
\end{deluxetable}

M31 clusters are reasonably well-resolved in \textit{HST}  imaging, but the effects of the
point spread function  (PSF) are not completely negligible in fitting surface brightness profiles.
A typical  cluster core radius ($\sim0.5$~pc) subtends 0\farcs13 at the distance of M31,
only slightly larger than the HST PSF. {\sc gridfit} accounts for
the PSF by convolving model profiles with the PSF before comparison
to the data. PSF profiles  for each specific instrument and filter combination were generated using the 
Web interface to the TinyTim PSF simulator \citep{krist06}.
For all instrument/filter 
combinations, a blackbody source with temperature 4000 K and the nominal telescope focus were assumed. 
For WFPC2 observations, we generated PSFs for individual cluster, corresponding with the 
cluster's position on the detector. For ACS observations, we assumed the clusters to be 
in the center of the chip 1 detector. The different treatment for the two cameras is consistent with
previous work \citep{barmby02,barmby07} and with the observation that
the ACS spatial PSF variation is small compared to that in WFPC2 \citep{sirianni05}.
We consider modeling the ACS PSF and data reduction at a finer level of detail to be beyond the scope of
this work, and so caution that our derived cluster core parameters are subject to non-negligable
uncertainties which are not easily quantified.

\subsection{Model-fitting and Results}
\label{sec:modfit}

We fit four models to the surface brightness profiles; these are described in detail by \citet{mclaughlin05}.
The \citet[hereafter referred to as K66]{king66} model is the `standard' model used when describing GCs and is 
characterized by a single-mass, isotropic, isothermal sphere. The \citet{wilson75} model is a slight modification of 
the \citetalias{king66} models with an extra term in the distribution function that causes Wilson models to be more spatially extended. 
The \citet{king62} model is an analytical parametrization of the surface brightness profile  sometimes
used in studies of marginally-resolved clusters and
the `power law  with core' model of \citet{elson87} is often used to describe young clusters.
Table~\ref{tab:modfits} gives the fitting results for each cluster.

\begin{deluxetable}{lccrrrrrrrr}
\tabletypesize{\scriptsize}
\rotate
\tablewidth{0pt}
\tablecaption{Model-fitting results
\label{tab:modfits}}
\tablehead{
\colhead{Name} & \colhead{Camera/} &
\colhead{$N_{\rm pts}$} & \colhead{Model} & \colhead{$\chi_{\rm min}^2$}   &
\colhead{$I_{\rm bkg}$} & \colhead{$W_0$} & \colhead{$c$}         &
\colhead{$\mu_0$} & \colhead{$\log\,r_0$} & \colhead{$\log\,r_0$}         \\
\colhead{} & \colhead{Filter}  & \colhead{}   &
\colhead{} & \colhead{} & \colhead{[$L_\odot\,{\rm pc}^{-2}$]} & \colhead{}  &
\colhead{} & \colhead{[mag arcsec$^{-2}$]} & \colhead{[arcsec]}              &
\colhead{[pc]} }
\startdata
\object[SKHV 148]{B086-G148}  & WFC/F435   & $26$       & K66  & $321.95$  & $240.42\pm35.61$  & $7.50^{+0.20}_{-0.20}$  & $1.68^{+0.06}_{-0.06}$  & $14.64^{+0.01}_{-0.01}$  & $-0.709^{+0.009}_{-0.010}$  & $-0.130^{+0.009}_{-0.010}$ \\
          ~~        & ~~      & & W  & $482.40$  & $148.41\pm56.15$  & $7.90^{+0.30}_{-0.30}$  & $3.10^{+0.15}_{-0.22}$  & $14.65^{+0.01}_{-0.01}$  & $-0.698^{+0.012}_{-0.011}$  & $-0.119^{+0.012}_{-0.011}$ \\
          ~~        & ~~     & & PL  & $438.38$  & $161.08\pm10.61$       & ---  & $3.10^{+0.03}_{-0.00}$  & $14.64^{+0.00}_{-0.00}$  & $-0.706^{+0.008}_{-0.000}$  & $-0.127^{+0.008}_{-0.000}$ \\
          ~~        & ~~    & & K62  & $368.87$  & $180.93\pm68.68$       & ---  & $2.08^{+0.82}_{-0.30}$  & $14.64^{+0.01}_{-0.01}$  & $-0.725^{+0.010}_{-0.009}$  & $-0.145^{+0.010}_{-0.009}$ \\
\object[Bol D091]{B091D-D057}  & WFC/F555   & $28$       & K66  & $315.28$  & $93.88\pm4.27$  & $15.90^{+0.10}_{-0.70}$  & $3.56^{+0.02}_{-0.16}$  & $7.91^{+0.34}_{-0.05}$  & $-2.918^{+0.140}_{-0.020}$  & $-2.339^{+0.140}_{-0.020}$ \\

\enddata
\tablecomments{
Table \ref{tab:modfits} is available in its entirety in the electronic  edition of the Journal, and at the end of the arxiv version. 
A short extract from it is shown here, for guidance regarding its form and content. 
Column descriptions:
$\chi_{\rm min}^2$: unreduced $\chi^2$ of best-fitting model;
$I_{\rm bkg}$: model-fit background intensity;
$W_0$: model-fit central potential;
$c = \log(r_t/r_0)$: model-fit concentration ($I(r_t)=0$);
$\mu_0$: model-fit central surface brightness, extinction-corrected in the native  bandpass of the data;
$\log\,r_0$: model-fit scale radius. 
Uncertainties are 68\% confidence intervals.
}
\end{deluxetable}

\begin{figure}
\plotone{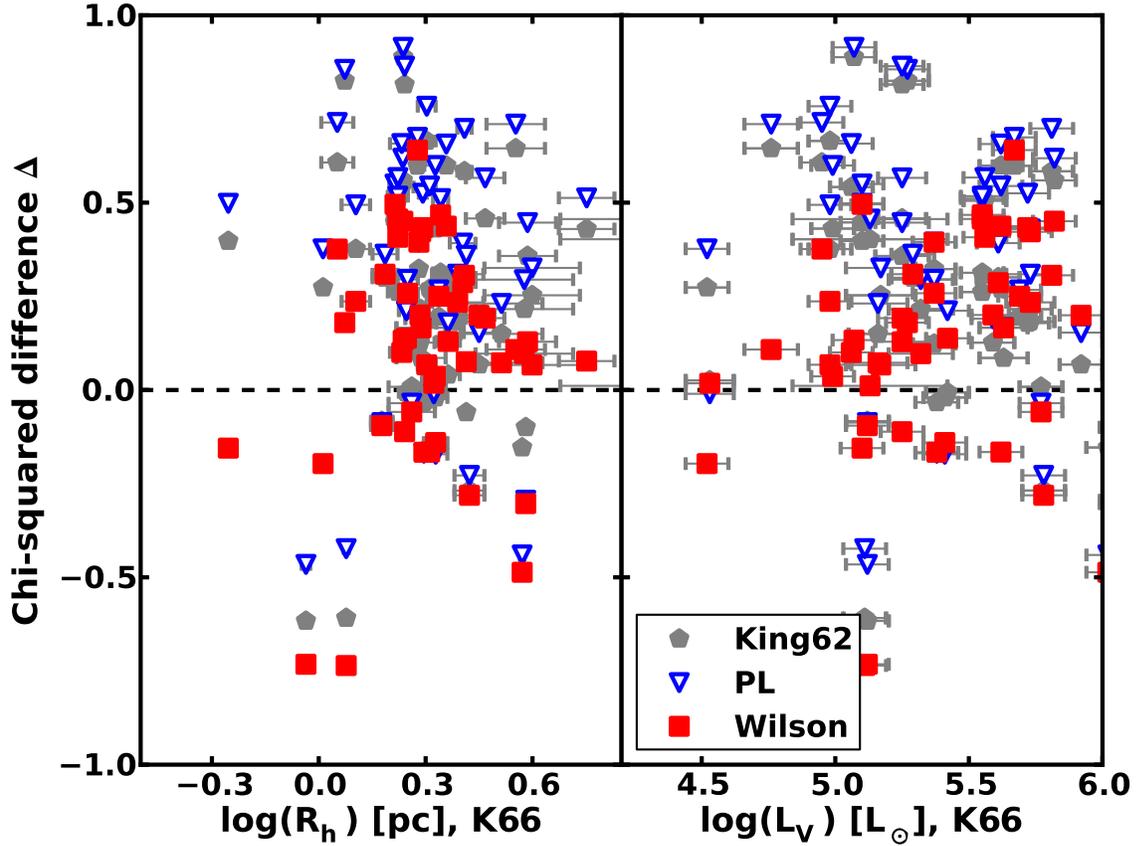}
\caption{Fitting $\chi^2$ difference  $\Delta$ (Eqn. \ref{eq: delta}) versus half-light radius (left)
and integrated $V$ band luminosity of the clusters (right), comparing \citetalias{king66}
models to \citet{wilson75}, power-law \citep{elson87} and \citet{king62}.
$\Delta$ is positive if the \citetalias{king66} model is a better fit than the other model. 
\label{fig:delta}}
\end{figure}

\begin{figure}
\includegraphics*[scale=0.5]{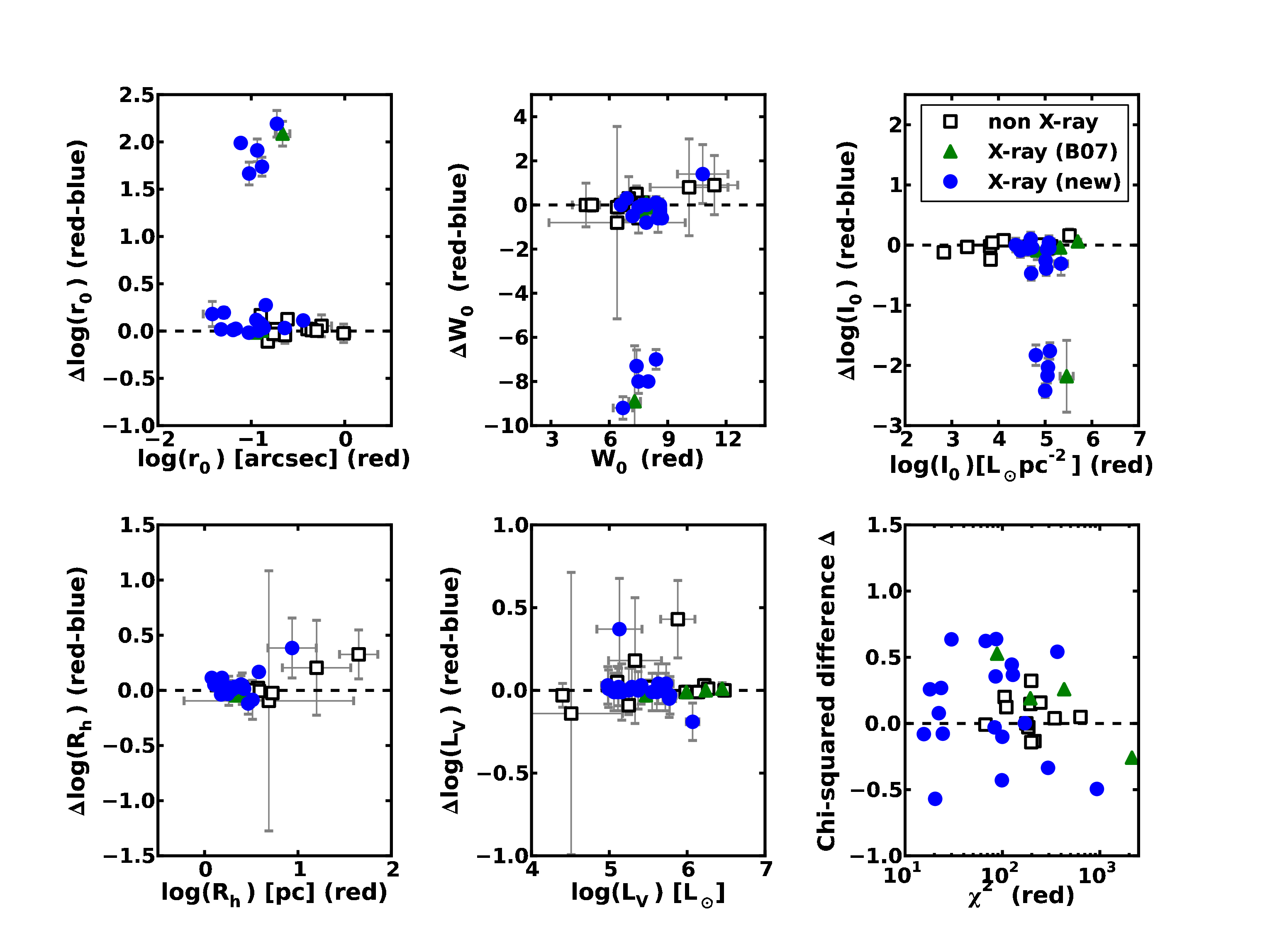}
\caption{
Comparison of surface-brightness profile fit results for the same objects observed in different
bandpasses. 
Top panels: scale radius $r_0$ (left), central potential $W_0$ (center), 
and central surface brightness converted to the $V$-band $I_0$ (right).
Bottom panels:  half-light radius $R_h$ (left), total luminosity $L_V$ (center), 
and $\chi^2$ difference  $\Delta$, defined such that $\Delta>0$ implies a better fit in the red band (right).
\label{fig:col_comp}}
\end{figure}

\begin{figure}
\plotone{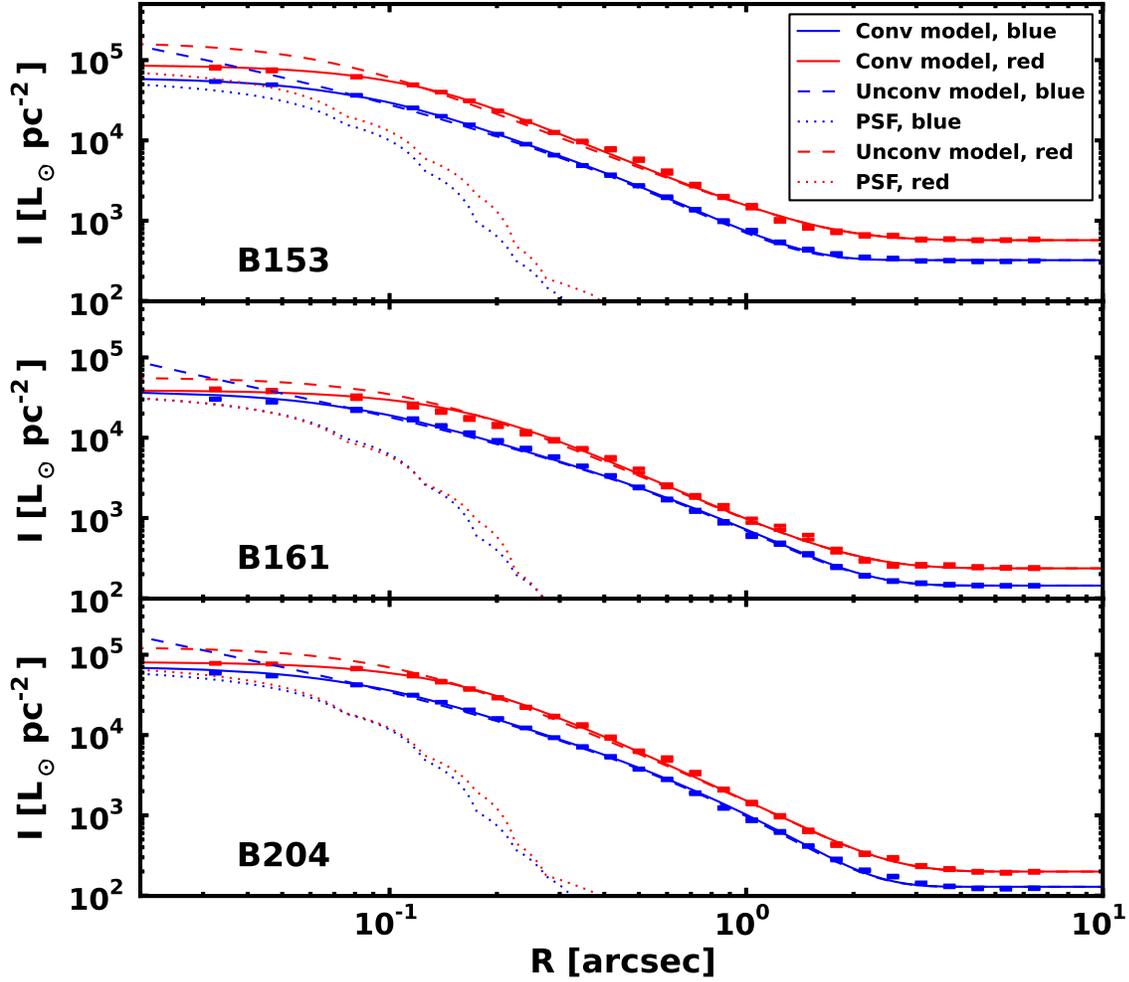}
\caption{
Surface brightness profile for  three clusters with different best-fit \citetalias{king66} in different bandpasses.
For each profile (points with error bars) the best-fitting model is overplotted, with both PSF-convolved (solid line)
and unconvolved (dashed line) versions indicated. The PSF is plotted as a
dotted line, with the peak intensity arbitrarily scaled to match the peak intensity
of the PSF-convolved model.
\label{fig:sbprofs}}
\end{figure}

In order to determine which model family we should use for further analysis, we want to determine which model is best. 
We do this by calculating the $\chi^2$ difference $\Delta$, as in \citet{mclaughlin05}. Each model fit has a value of $\Delta$ associated with it, defined as:
\begin{equation}
\Delta = \frac{\chi^2_{\rm model} - \chi^2_{\rm King66}}{\chi^2_{\rm model} + \chi^2_{\rm King66}}
\label{eq: delta}
\end{equation} 
According to the equation above, we compare each model fit with \citetalias{king66} model fits.
The value of $\Delta$ will be positive if the \citetalias{king66} model is a better fit and negative if the model being compared 
is better than the \citetalias{king66} model.
Figure~\ref{fig:delta} shows  $\Delta$ plotted against projected half-light radius
and  luminosity from  the \citetalias{king66} fit for each cluster/bandpass combination. 
There are no systematic trends whereby one model family is a better fit for clusters
in a specific luminosity or size range.
For the majority of clusters, we find a positive value of $\Delta$, meaning 
that the \citetalias{king66} model is a better fit than the alternatives. 
\citetalias{king66} models are also the most widely used in the literature, therefore further analysis 
of these GCs will be done using the \citetalias{king66} model fits.
Table~\ref{tab:phot} gives various derived  parameters for all \citetalias{king66} models for each cluster  
\citep[the details of their calculation are given by][]{mclaughlin08a}.

Comparing surface brightness profile fits of the same cluster in more than one bandpass is a useful check
on our model-fitting. Including both the B07 clusters and objects newly analyzed here, there are 35
clusters observed in more than one bandpass, with the redder bandpass usually but not always F814W.
Of these 35, 23 are LMXB-containing clusters.
The fitting results in multiple bands are compared in Figure~\ref{fig:col_comp}. The 
$\chi^2$ difference  $\Delta$ is defined such that $\Delta>0$ implies a better fit in the red band, and shows
that the model fits do not systematically prefer red or blue bandpasses. 
The agreement between fit parameters is good for most clusters, and the agreement in $L_V$ 
in particular implies that the conversion to the $V$-band discussed at the end of Section~\ref{sec:sample} 
works well. There are several clusters for which the model fits return a high concentration
and small scale radius  in the blue bandpass and a more typical set of parameters in the red filter.
Some sample surface brightness profiles and model fits for these clusters are shown in Figure~\ref{fig:sbprofs}.
The agreement between rather different models and similar profiles illustrates that there is some degeneracy in the fits;
many of the clusters with small blue $r_0$ values show double minima
in $\chi^2$  as a function of $W_0$, with the second minimum at a smaller $W_0$/larger $r_0$.
This is a reminder that even the spatial resolution of \textit{HST} may not be sufficient to resolve the
cores of all M31 star clusters. 

A closer look at Figure~\ref{fig:col_comp} shows that the non-outlier clusters also show  slight
systematic offsets between bandpasses: the central potential is higher (median $\Delta W_0=0.1$) and scale radius
lower (median $\Delta \log r_0=-0.03$) when measured in the bluer bandpass.  These offsets
are larger (median $\Delta W_0=0.3$, median $\Delta \log r_0=-0.09$) when only the LMXB clusters are considered.
Bluer cores for LMXB clusters would be intriguing in light of the finding by \citet{peacock11a} 
that M31 GCs with higher central densities have bluer ultraviolet colors.
(Those authors suggest that a population of extreme horizontal branch stars
formed through dynamical interactions could account for this effect.)
We carried out several tests to determine whether this offset in profile scale radius was the result of measurement bias,
and found that the most obvious differences between red and blue filters---PSF and background levels---were not
responsible for the offset. However, given the degeneracies in fits noted above, we are reluctant to claim too much significance
for this finding. A larger sample of clusters with more detailed, position-dependent PSF modeling
should provide a better understanding of this effect.

\begin{figure}
\plotone{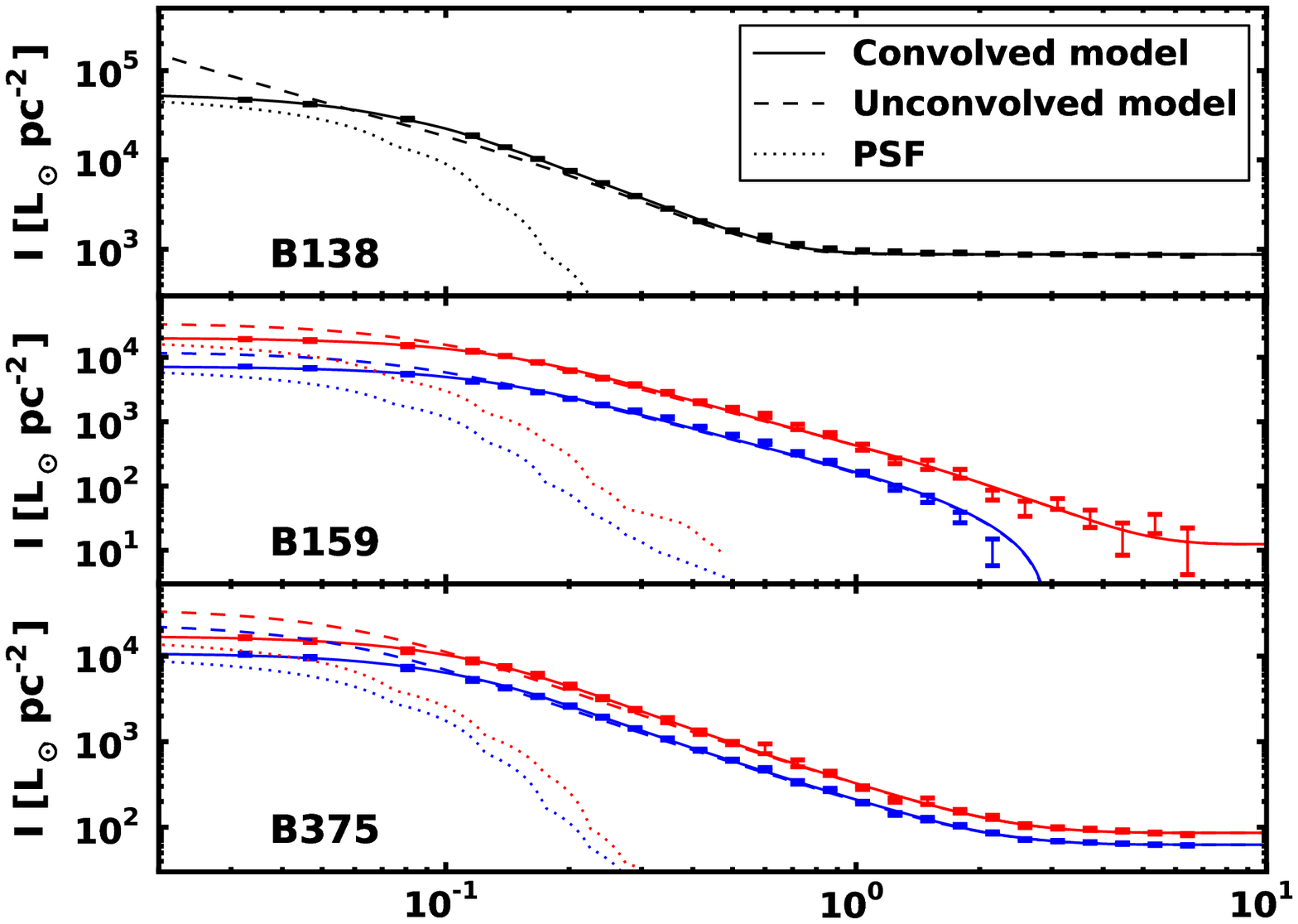}
\caption{
Surface brightness profile for  three clusters discussed in the text.
For each cluster the best-fitting \citetalias{king66} model is overplotted, with both PSF-convolved (solid line)
and unconvolved (dashed line) versions indicated. The PSF is plotted as a
dashed line, with the peak intensity arbitrarily scaled to match the peak intensity
of the PSF-convolved model.
Note that the vertical scales of the three panels are different.
\label{fig:sbprofs2}}
\end{figure}

\begin{deluxetable}{llrrrrrrr}
\tablewidth{0pt}
\tabletypesize{\tiny}
\tablecaption{Derived photometric and mass-related parameters from K66 model-fitting
\label{tab:phot}}
\tablehead{
\colhead{Name} &
\colhead{Band} &
\colhead{$\log\,R_c$} &
\colhead{$\log\,R_h$} &
\colhead{$\log\,L_{V}$}  &
\colhead{$\log\,M_{\rm tot}$} &
\colhead{$\log\,\rho_0$} & 
\colhead{$\log\,\Gamma$} & 
\colhead{$\log\,\Gamma_h$} 
\\ 
\colhead{} &
\colhead{} &
\colhead{[pc]} & 
\colhead{[pc]} &
\colhead{[$L_{\odot, V}$]} &
\colhead{[$M_\odot$]} &
\colhead{[$M_\odot$ pc$^{-3}$]} &
\colhead{} &
\colhead{}
}
\startdata
\object[SKHV 148]{B086-G148}                &  F435* & $-0.15\pm0.01$& $ 0.45\pm0.04$& $ 5.92\pm0.08$& $ 6.19\pm0.09$& $ 5.09\pm0.08$& $ 7.34\pm0.05$& $ 8.38\pm0.06$\\
\object[Bol D091]{B091D-D057}               &  F555 & $-2.34\pm0.14$& $ 0.24\pm0.00$& $ 5.82\pm0.08$& $ 6.21\pm0.08$& $ 9.84\pm0.09$& $10.09\pm0.09$& $ 8.84\pm0.05$\\
\object[Bol D091]{B091D-D057}               &  F814* & $-0.17\pm0.01$& $ 0.26\pm0.07$& $ 5.77\pm0.08$& $ 6.16\pm0.09$& $ 5.26\pm0.09$& $ 7.55\pm0.06$& $ 8.72\pm0.07$\\
\object[2MASXi J0042250+405717]{B094-G156}  &  F555* & $-0.33\pm0.01$& $ 0.34\pm0.02$& $ 5.55\pm0.08$& $ 6.02\pm0.08$& $ 5.41\pm0.08$& $ 7.45\pm0.05$& $ 8.35\pm0.05$\\
\object[NBol 002]{B096-G158}                &  F814* & $-0.17\pm0.01$& $ 0.25\pm0.02$& $ 5.42\pm0.08$& $ 5.91\pm0.08$& $ 5.02\pm0.08$& $ 7.19\pm0.05$& $ 8.37\pm0.05$\\
\object[NBol 001]{B107-G169}                &  F475 & $-0.37\pm0.02$& $ 0.34\pm0.03$& $ 5.69\pm0.08$& $ 6.02\pm0.08$& $ 5.50\pm0.10$& $ 7.52\pm0.07$& $ 8.36\pm0.06$\\
\object[NBol 001]{B107-G169}                &  F814* & $-0.31\pm0.01$& $ 0.39\pm0.07$& $ 5.73\pm0.09$& $ 6.07\pm0.09$& $ 5.39\pm0.09$& $ 7.46\pm0.06$& $ 8.33\pm0.07$\\
\object[SKHV 178]{B110-G172}                &  F606* & $-0.19\pm0.02$& $ 0.36\pm0.03$& $ 6.08\pm0.08$& $ 6.48\pm0.08$& $ 5.53\pm0.09$& $ 7.92\pm0.06$& $ 8.99\pm0.05$\\
\object[SKHV 178]{B116-G178}                &  F555* & $-0.34\pm0.02$& $ 0.29\pm0.02$& $ 5.73\pm0.08$& $ 6.13\pm0.08$& $ 5.57\pm0.08$& $ 7.69\pm0.05$& $ 8.62\pm0.05$\\
\object[SKHV 178]{B116-G178}                &  F814 & $-0.30\pm0.01$& $ 0.29\pm0.01$& $ 5.72\pm0.08$& $ 6.12\pm0.08$& $ 5.49\pm0.09$& $ 7.63\pm0.06$& $ 8.60\pm0.05$\\
\object[SKHV 176]{B117-G176}                &  F336* & $-0.34\pm0.02$& $ 0.75\pm0.08$& $ 6.36\pm0.10$& $ 6.63\pm0.10$& $ 5.75\pm0.09$& $ 7.95\pm0.06$& $ 8.44\pm0.07$\\
\object[SKHV 187]{B128-G187}                &  F555* & $-0.43\pm0.01$& $ 0.21\pm0.02$& $ 5.10\pm0.08$& $ 5.52\pm0.08$& $ 5.24\pm0.08$& $ 6.99\pm0.05$& $ 7.85\pm0.05$\\
\object[SKHV 187]{B128-G187}                &  F814 & $-0.35\pm0.01$& $ 0.18\pm0.03$& $ 5.12\pm0.08$& $ 5.54\pm0.08$& $ 5.09\pm0.09$& $ 6.93\pm0.06$& $ 7.96\pm0.05$\\
\object[SKHV 192]{B135-G192}                &  F555 & $-2.11\pm0.12$& $ 0.36\pm0.00$& $ 5.62\pm0.08$& $ 5.89\pm0.08$& $ 8.94\pm0.18$& $ 9.19\pm0.13$& $ 8.12\pm0.05$\\
\object[SKHV 192]{B135-G192}                &  F814* & $-0.45\pm0.03$& $ 0.40\pm0.04$& $ 5.61\pm0.08$& $ 5.89\pm0.08$& $ 5.52\pm0.10$& $ 7.37\pm0.07$& $ 8.03\pm0.06$\\
\object[C401141021]{B138}                   &  F475* & $-1.72\pm0.05$& $-0.25\pm0.00$& $ 5.10\pm0.08$& $ 5.67\pm0.08$& $ 8.60\pm0.17$& $ 9.47\pm0.12$& $ 9.01\pm0.05$\\
\object[C401540596]{B144}                   &  F475 & $-0.92\pm0.04$& $ 0.24\pm0.01$& $ 5.07\pm0.08$& $ 5.67\pm0.08$& $ 6.48\pm0.09$& $ 7.89\pm0.06$& $ 8.03\pm0.05$\\
\object[C401540596]{B144}                   &  F814* & $-0.72\pm0.03$& $ 0.23\pm0.03$& $ 5.06\pm0.08$& $ 5.65\pm0.08$& $ 6.03\pm0.11$& $ 7.60\pm0.07$& $ 8.01\pm0.05$\\
\object[C401840589]{B146}                   &  F475 & $-0.49\pm0.03$& $ 0.60\pm0.13$& $ 5.17\pm0.13$& $ 5.50\pm0.13$& $ 5.08\pm0.09$& $ 6.64\pm0.06$& $ 7.05\pm0.10$\\
\object[C401840589]{B146}                   &  F814* & $-0.38\pm0.02$& $ 0.51\pm0.12$& $ 5.16\pm0.11$& $ 5.49\pm0.11$& $ 4.87\pm0.09$& $ 6.55\pm0.06$& $ 7.21\pm0.09$\\
\object[SKHV 200]{B148-G200}                &  F475 & $-0.36\pm0.02$& $ 0.28\pm0.04$& $ 5.59\pm0.08$& $ 5.90\pm0.09$& $ 5.41\pm0.09$& $ 7.39\pm0.06$& $ 8.28\pm0.06$\\
\object[SKHV 200]{B148-G200}                &  F814* & $-0.36\pm0.01$& $ 0.29\pm0.07$& $ 5.63\pm0.09$& $ 5.95\pm0.09$& $ 5.45\pm0.09$& $ 7.46\pm0.06$& $ 8.35\pm0.07$\\
\object[SKHV 203]{B150-G203}                &  F555* & $-0.12\pm0.01$& $ 0.29\pm0.02$& $ 5.38\pm0.08$& $ 5.87\pm0.08$& $ 4.83\pm0.09$& $ 7.00\pm0.06$& $ 8.21\pm0.05$\\
\object[SKHV 203]{B150-G203}                &  F814 & $-0.09\pm0.02$& $ 0.33\pm0.03$& $ 5.41\pm0.08$& $ 5.90\pm0.08$& $ 4.76\pm0.09$& $ 6.96\pm0.06$& $ 8.19\pm0.05$\\
\object[Bol 153]{B153}                      &  F475 & $-2.52\pm0.02$& $ 0.07\pm0.00$& $ 5.27\pm0.08$& $ 5.77\pm0.08$& $ 9.93\pm0.08$& $ 9.85\pm0.05$& $ 8.51\pm0.05$\\
\object[Bol 153]{B153}                      &  F814* & $-0.55\pm0.03$& $ 0.19\pm0.04$& $ 5.29\pm0.08$& $ 5.78\pm0.08$& $ 5.78\pm0.10$& $ 7.58\pm0.07$& $ 8.30\pm0.06$\\
\object[Bol 159]{B159}                      &  F435 & $-0.38\pm0.03$& $ 0.58\pm0.14$& $ 5.32\pm0.12$& $ 5.63\pm0.12$& $ 4.97\pm0.10$& $ 6.70\pm0.07$& $ 7.29\pm0.10$\\
\object[Bol 159]{B159}                      &  F475 & $-0.44\pm0.03$& $ 0.59\pm0.08$& $ 5.25\pm0.10$& $ 5.56\pm0.10$& $ 5.03\pm0.09$& $ 6.67\pm0.06$& $ 7.17\pm0.07$\\
\object[Bol 159]{B159}                      &  F814* & $-0.46\pm0.01$& $ 0.47\pm0.05$& $ 5.25\pm0.09$& $ 5.55\pm0.09$& $ 5.15\pm0.09$& $ 6.81\pm0.06$& $ 7.39\pm0.06$\\
\object[SKHV 215]{B161-G215}                &  F475 & $-2.04\pm0.10$& $ 0.28\pm0.01$& $ 5.37\pm0.08$& $ 5.69\pm0.08$& $ 8.69\pm0.28$& $ 8.95\pm0.19$& $ 7.97\pm0.05$\\
\object[SKHV 215]{B161-G215}                &  F814* & $-0.32\pm0.02$& $ 0.25\pm0.03$& $ 5.37\pm0.08$& $ 5.69\pm0.08$& $ 5.14\pm0.09$& $ 7.06\pm0.06$& $ 8.04\pm0.05$\\
\object[SKHV 217]{B163-G217}                &  F555 & $-1.03\pm0.09$& $-0.04\pm0.01$& $ 5.12\pm0.08$& $ 5.66\pm0.08$& $ 6.92\pm0.20$& $ 8.33\pm0.14$& $ 8.56\pm0.05$\\
\object[SKHV 217]{B163-G217}                &  F814* & $-0.85\pm0.10$& $ 0.08\pm0.01$& $ 5.11\pm0.08$& $ 5.66\pm0.08$& $ 6.43\pm0.22$& $ 7.95\pm0.15$& $ 8.34\pm0.05$\\
\object[Bol 164]{B164-V253}                 &  F475 & $-0.77\pm0.01$& $ 0.05\pm0.04$& $ 4.95\pm0.08$& $ 5.44\pm0.09$& $ 6.06\pm0.09$& $ 7.54\pm0.06$& $ 8.06\pm0.06$\\
\object[Bol 164]{B164-V253}                 &  F814* & $-0.75\pm0.02$& $ 0.10\pm0.04$& $ 4.98\pm0.08$& $ 5.47\pm0.08$& $ 6.00\pm0.10$& $ 7.49\pm0.07$& $ 8.00\pm0.06$\\
\object[SKHV 233]{B182-G233}                &  F555 & $-0.55\pm0.01$& $ 0.41\pm0.02$& $ 5.81\pm0.08$& $ 6.13\pm0.08$& $ 5.98\pm0.09$& $ 7.87\pm0.06$& $ 8.38\pm0.05$\\
\object[SKHV 233]{B182-G233}                &  F814* & $-0.28\pm0.01$& $ 0.42\pm0.04$& $ 5.78\pm0.08$& $ 6.10\pm0.08$& $ 5.32\pm0.09$& $ 7.42\pm0.06$& $ 8.30\pm0.06$\\
\object[SKHV 235]{B185-G235}                &  F435* & $-0.43\pm0.01$& $ 0.28\pm0.01$& $ 5.67\pm0.08$& $ 6.08\pm0.08$& $ 5.74\pm0.09$& $ 7.76\pm0.06$& $ 8.57\pm0.05$\\
\object[2MASX J00434551+4136578]{B193-G244} &  F475* & $-0.61\pm0.03$& $ 0.31\pm0.02$& $ 5.62\pm0.08$& $ 6.18\pm0.08$& $ 6.24\pm0.09$& $ 8.13\pm0.06$& $ 8.65\pm0.05$\\
\object[SKHV 254]{B204-G254}                &  F475 & $-2.27\pm0.12$& $ 0.22\pm0.01$& $ 5.56\pm0.08$& $ 5.95\pm0.08$& $ 9.45\pm0.22$& $ 9.64\pm0.16$& $ 8.48\pm0.05$\\
\object[SKHV 254]{B204-G254}                &  F814* & $-0.37\pm0.01$& $ 0.22\pm0.03$& $ 5.55\pm0.08$& $ 5.94\pm0.08$& $ 5.51\pm0.09$& $ 7.52\pm0.06$& $ 8.47\pm0.05$\\
\object[SKHV 264]{B213-G264}                &  F435* & $-0.78\pm0.05$& $ 0.24\pm0.00$& $ 5.25\pm0.08$& $ 5.62\pm0.08$& $ 6.13\pm0.10$& $ 7.63\pm0.07$& $ 7.95\pm0.05$\\
\object[SKHV 011]{B293-G011}                &  F555* & $-0.00\pm0.01$& $ 0.41\pm0.01$& $ 6.26\pm0.08$& $ 6.53\pm0.08$& $ 5.14\pm0.08$& $ 7.70\pm0.05$& $ 8.97\pm0.05$\\
\object[SKHV 011]{B293-G011}                &  F606 & $ 0.09\pm0.03$& $ 0.57\pm0.01$& $ 6.02\pm0.08$& $ 6.29\pm0.08$& $ 4.58\pm0.09$& $ 7.04\pm0.06$& $ 8.29\pm0.05$\\
\object[SKHV 011]{B293-G011}                &  F814 & $ 0.11\pm0.01$& $ 0.58\pm0.01$& $ 6.07\pm0.08$& $ 6.34\pm0.08$& $ 4.56\pm0.09$& $ 7.07\pm0.06$& $ 8.35\pm0.05$\\
\object[SKHV 307]{B375-G307}                &  F475 & $-0.62\pm0.02$& $ 0.30\pm0.03$& $ 4.98\pm0.08$& $ 5.33\pm0.08$& $ 5.42\pm0.12$& $ 6.89\pm0.08$& $ 7.39\pm0.05$\\
\object[SKHV 307]{B375-G307}                &  F814* & $-0.60\pm0.02$& $ 0.33\pm0.03$& $ 4.99\pm0.08$& $ 5.34\pm0.08$& $ 5.36\pm0.11$& $ 6.85\pm0.07$& $ 7.36\pm0.05$\\
\object[CXO J004246.0+411736]{BH16}         &  F435* & $-1.11\pm0.06$& $ 0.01\pm0.00$& $ 4.52\pm0.08$& $ 4.85\pm0.08$& $ 6.26\pm0.23$& $ 7.17\pm0.16$& $ 7.25\pm0.05$\\
\object[NBol 021]{NB21}                     &  F435 & $ 0.28\pm0.00$& $ 0.32\pm0.03$& $ 4.53\pm0.09$& $ 4.84\pm0.09$& $ 2.97\pm0.09$& $ 5.01\pm0.06$& $ 6.61\pm0.06$\\
\object[NBol 021]{NB21}                     &  F475* & $-0.63\pm0.01$& $ 0.55\pm0.08$& $ 4.76\pm0.10$& $ 5.08\pm0.10$& $ 5.01\pm0.08$& $ 6.25\pm0.05$& $ 6.51\pm0.07$\\
\object[NBol 021]{NB21}                     &  F814 & $-0.62\pm0.02$& $ 0.94\pm0.26$& $ 5.13\pm0.29$& $ 5.45\pm0.29$& $ 5.00\pm0.08$& $ 6.26\pm0.05$& $ 6.30\pm0.22$\\
\enddata
\tablecomments{
Asterisks indicate the model fit used for subsequent analysis.
Column descriptions:
$R_c$, the model projected  core radius, at which intensity is half the central value;
$R_h$, the model projected half-light, or effective, radius (that contains half the total luminosity in projection);
$L_V$, the total integrated model luminosity in the $V$ band;
$M_{\rm tot}= L_V (M/L)_V$, the integrated model mass; 
${\rho}_0$: the central volume density;
$\Gamma=\rho_0^{3/2} R_c^2$, the stellar collision rate computed in the core; and
$\Gamma_h=\rho_h^{3/2} R_h^2=M_{\rm tot}^{3/2}R_h^{-5/2}$, the stellar collision rate computed at half-light radius.
}
\end{deluxetable}

The model fits for a few clusters warrant specific comments: Figure~\ref{fig:sbprofs2} shows 
the surface brightness profile for each. Cluster B138 has a surface brightness profile only in the F475W
band as the F814W data were saturated. Its best-fit model has a high central density and small scale radius,
like the blue-versus-red outliers discussed above. Cluster BH16 has a similar profile. 
Clusters B159 and B375-G307 are examples of two clusters where,
unlike the clusters shown in Figure~\ref{fig:sbprofs}, the red and blue band model fit parameters are not drastically different.
\citet{strader11} found cluster B159 to have an unresolved core, but our fitting procedure finds it to be rather
unremarkable. 
B375 is host to the most X--ray luminous GC in M31 \citep{tp04,distefano02} but 
its surface brightness profile is not particularly unusual and in fact is close to that of B159.

Given the heterogeneous set of archival data we have to work with,
a systematic difference in cluster properties between bands raises the important question of
how best to compile a dataset for comparison of structural properties between LMXB and non-LMXB clusters.
While overall cluster properties such as half-light radius and total luminosity are not extremely sensitive
to the details of surface-brightness model fitting (see the bottom panel of Figure~\ref{fig:col_comp}), 
the same is not true for core parameters such as central density and core radius---precisely those which 
are used to compute the collision rate.%
\footnote{In more distant galaxies, only parameters relating to the
half-light radius can be measured, and  \citet{sivakoff07} used these to compute the collision rate as $\Gamma_h$.
However, \citet{maccarone11} concluded that ``it is dangerous to try to infer physical information about how X-ray binaries 
are formed by using $\Gamma_h$.'' } Using measurements from a single bandpass would mitigate
any measurement bias and/or systematic offsets between bands and also any issues involved in transforming 
measurements to a single band, but at the cost of greatly reduced sample size. For example,
24 LMXB clusters have observations in F814W, but only 12 non-LMXB clusters do; the numbers are
reversed for F606W. While a comprehensive comparison of LMXB and non-LMXB clusters in the
northern half of M31 will be possible after the completion of the PHAT survey, for now 
we proceed with the analysis by choosing one fit per object on the basis of the lowest  $\chi^2$
(for the `outlier' clusters in Figure~\ref{fig:col_comp}, we use the red bandpass fit).
The chosen bandpass for each LMXB object is indicated in Table~\ref{tab:phot}.

\section{Analysis and discussion}
\label{sec:disc}

\begin{figure}
\plotone{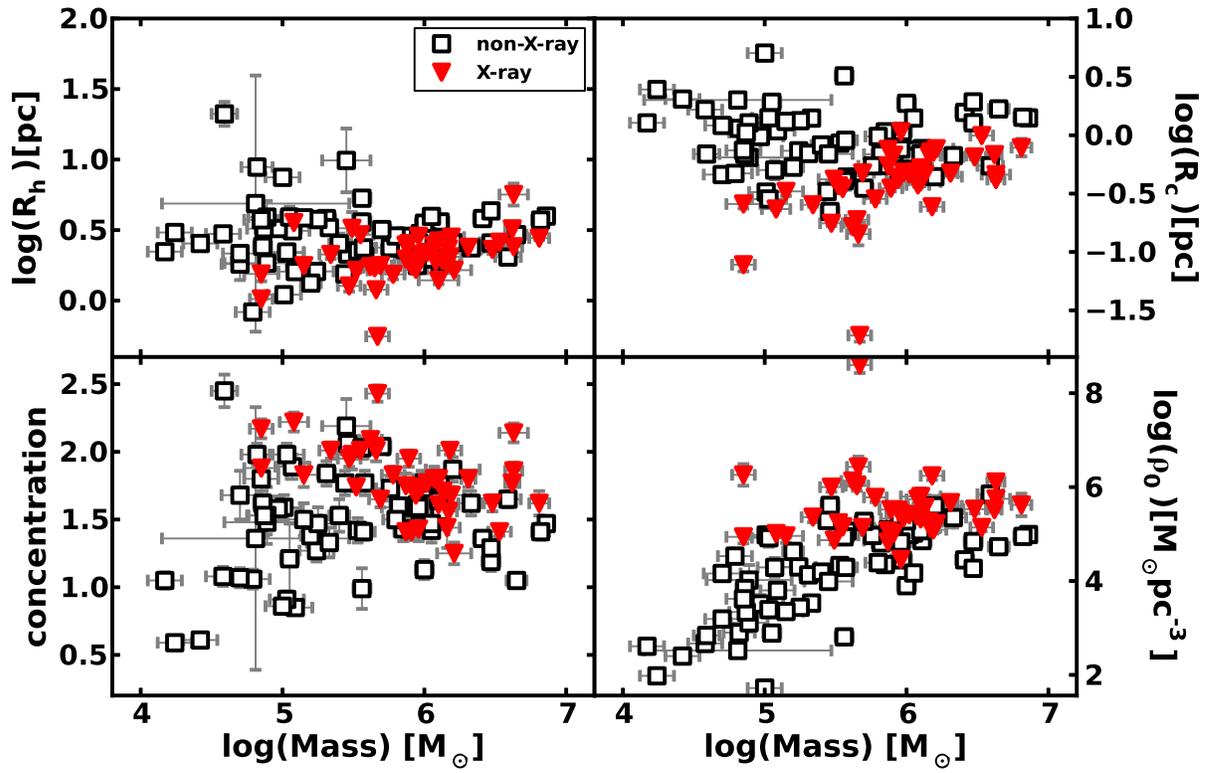}
\caption{
Cluster structural parameters as a function of mass for M31 GCs with and without LMXBs.
The two LMXB clusters with small core radii and high central densities are B138 and BH16,
discussed in Section~\ref{sec:modfit} of the text.
\label{fig:struct_compare}}
\end{figure}

Figure~\ref{fig:struct_compare} shows the distribution of \citetalias{king66} structural parameters
for M31 clusters  with and without LMXBs. As in previous work
\citep[e.g.][]{tp04,jordan07,peacock09}, we find that GCs hosting  LMXBs are more
massive than typical clusters.
Also consistent with previous work \citep[e.g.][]{mclaughlin05,barmby07,masters10}, 
we find that,  for the overall cluster population, the half-light radius, core radius, and concentration are not strongly correlated with 
cluster mass. However, compared to non-LMXB clusters of the same mass,
LMXB clusters have larger median concentrations and central densities,
and smaller radii.   We compare the two sets of clusters with a 
Kolmogorov-Smirnov (KS) test, we find that the distributions of all of the parameters
plotted in Figure~\ref{fig:struct_compare}, as well as those in Figure~\ref{fig:x_compare}, differ between the LMXB and non-LMXB samples at the $p<0.01$ level.
We remind the reader that these differences may well be due at least in part to the different
spatial distributions of the two samples (see Figure~\ref{fig:x_compare}).

Cluster core radius $\rho_0$ and central density $R_c$ both contribute to the stellar collision rate
$\Gamma$, for which the most commonly-used estimate is
\begin{equation}
\Gamma = \rho_0^{3/2}R_c^2
\label{eq:gamma}
\end{equation}
This theoretical expression for collision rate was first 
suggested by \citet{verbunt87} and arises by integrating the stellar collision rate per volume over the 
entire volume of the cluster (see \citealt{verbunt08} for a full derivation).
\citet{maccarone11} made the important point that  this expression ``implicitly assumes that the stellar velocity dispersion 
in the cluster core can be estimated correctly from the virial theorem,'' an assumption that we have
also made in this analysis.
While there are a number of different proxies for stellar collision rate  in the literature, 
\citet{maccarone11} found the expression above to be an accurate representation of the collision 
rate to within $\sim25$\% for most clusters.

\begin{figure}
\plotone{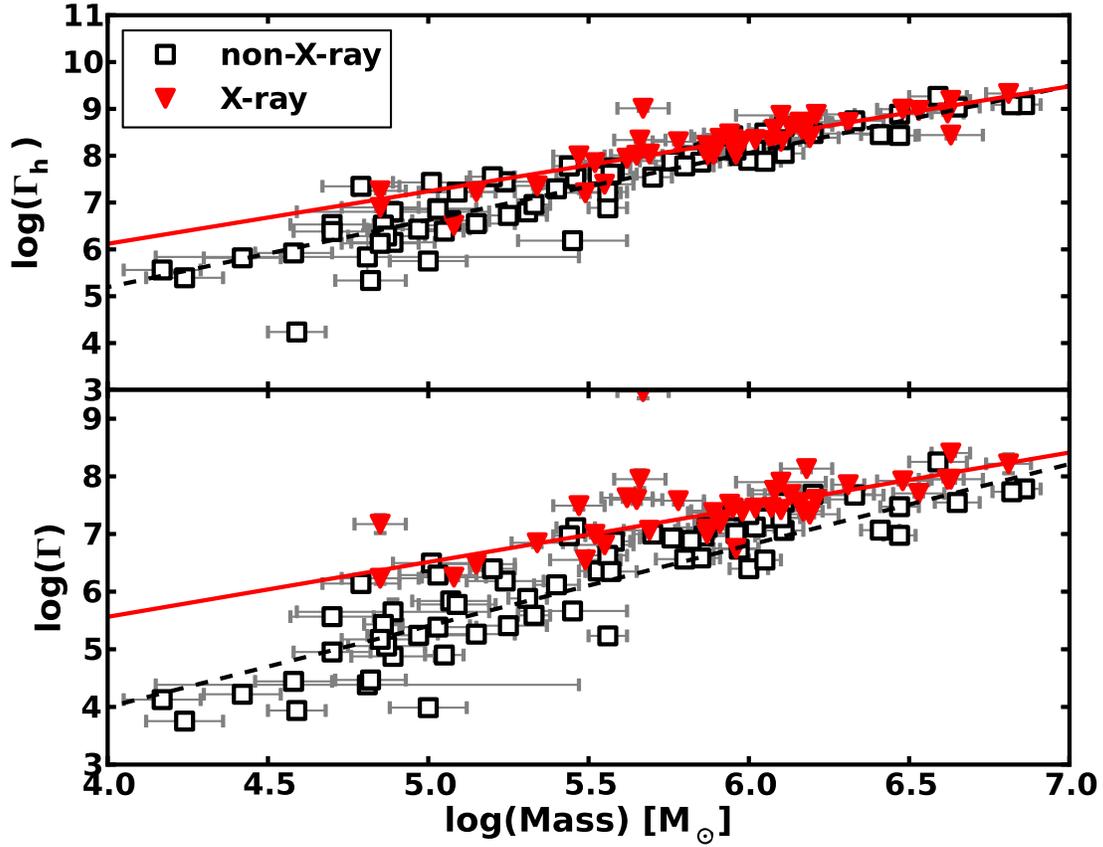}
\caption{
Two proxies for stellar collision rate versus cluster mass. (Bottom): Core proxy $\Gamma= \rho_0^{3/2} r_c^2$, 
(Top): Half-light proxy  $\Gamma_h= \rho_h^{3/2} r_h^2$.
Clusters with LMXBs typically show larger than expected stellar collision rates for their mass, with the
offset in $\Gamma$ larger than that for $\Gamma_h$, and larger at lower mass.
\label{fig:scr_mass}}
\end{figure}

The dynamical formation scenario for LMXBs implies that a large stellar 
collision rate should indicate a higher likelihood of the presence of an LMXB.
We computed  $\Gamma$ for the cluster sample using 
Equation~\ref{eq:gamma}, and find significant differences between clusters with and without LMXBs.
The LMXB-hosting clusters have  a larger average value of  $\Gamma$ than the non-LMXB clusters and
a KS test rejects the hypothesis that the two sets of  $\Gamma$ values are drawn from the same distribution at the $p<0.001$ level. 
$\Gamma$ is known to be correlated with mass, but LMXB clusters have
higher $\Gamma$ than non-LMXB clusters of the same mass, as shown in Figure~\ref{fig:scr_mass}.
This is consistent with the previous findings from \citetalias{peacock10}, and with the dynamical formation
theories of LMXBs. 
We also computed the collision rate at the half-light radius, $\Gamma_h$. We follow
\citet{sivakoff07} in defining $\Gamma_h = \rho_h^{3/2} R_h^2 = M^{3/2} R_h^{-5/2}$
(omitting the numerical factors). The top panel of Figure~\ref{fig:scr_mass} shows that 
there is less offset between the LMXB and non-LMXB populations in $\Gamma_h$
compared to $\Gamma$, particularly for high masses.  As for $\Gamma$, a KS test 
rejects the hypothesis that the two samples are drawn from the same distribution  at the $p<0.001$  level.

To what degree is the stellar collision rate enhanced in LMXB-containing clusters?
A simple way to estimate this is by looking at the offset in $\Gamma$ between LMXB
and non-LMXB clusters of the same mass. A weighted linear fit of the non-LMXB clusters
(dashed line in Figure~\ref{fig:scr_mass}) shows a relationship of $\Gamma \propto M_{\rm tot}^{1.4}$, slightly
lower than theoretical predictions by \citet{davies04}  and observational results by \citetalias{peacock10}.
A weighted linear fit of the complete sample of clusters with LMXBs (solid line) reveals a 
slightly shallower slope, suggesting that the stellar collision rate 
enhancement in clusters with LMXBs tends to be larger at smaller masses. 
In order to quantify  this enhancement, the predicted stellar collision rate from
the non-LMXB cluster fit was calculated for all clusters. 
The offset found for those  clusters with LMXBs was $\log \Gamma_{\rm X} = \log \Gamma_{\rm noX, pred}+ 0.78$. 
The stellar collision rate for a cluster with an LMXB is, on average, $\sim6$ times larger than 
that of a similar-mass cluster without an LMXB. The corresponding offset for
$\Gamma_h$ is $\Delta (\log \Gamma_h) = 0.312$, or a multiplicative factor of 2.1.

We have shown that, compared to non-LMXB clusters, LMXB-containing globular clusters in
M31 have enhanced stellar collision rates $\Gamma$. However, there are at least two possible contaminating
factors involved: metallicity and galactocentric distance.
Previous work has suggested that LMXBs tend to reside in clusters with higher metallicity,
both in the Milky Way and M31 \citep{bellazzini95} and in the elliptical galaxy NGC 4472 \citep{kundu02};
the latter authors  found that red metal-rich clusters are 3 times more likely to host an LMXB than blue metal-poor clusters. 
Average M31 cluster metallicity declines with galactocentric distance \citep{fan08},  while average
cluster size increases \citep{barmby07}, and of course size is an important contributor to $\Gamma$.
The other contributor to  $\Gamma$, central density, is strongly related to mass (Figure~\ref{fig:struct_compare}).
\citetalias{peacock10} plotted the residual from the stellar collision rate versus mass relationship 
against metallicity and inferred that metallicity could not explain the offset seen in collision rate for LMXB clusters 
\citepalias[see Figure 8 in][]{peacock10}. 
We have  a cluster sample nearly three times the size of that used by \citetalias{peacock10}, but
one in which the LMXB and non-LMXB sub-samples differ in metallicity and location (see Figure~\ref{fig:x_compare}),
so it is important to attempt to control for these differences.

One approach to analyzing the relationship between metallicity, location, structure, and LMXB occurrence
in globular clusters involves the use of logistic regression. This technique, not commonly used in astronomy,
treats the problem of modeling a binary response variable. The probability of success $p$ is transformed
by the logit function $\ln [p/(1-p)]$, which is then modeled as a linear function of the independent variables
using regression techniques. An important characteristic of logistic regression is the fact that the 
resulting parameters are not biased by the use of retrospective sampling (i.e., of sets of `successful'
and `unsuccessful' cases) as opposed to random, prospective sampling \citep{abraham06}. Our sample
selection more closely matches the retrospective approach.
We used maximum likelihood estimation as implemented in the Python {\tt statsmodels} package
to model the probability of a cluster containing an LMXB; the results of the logistic regression analysis
are given in Table~\ref{tab:logit}.  Experiments showed that the fit results were insensitive to the inclusion of the clusters B138 and BH16, 
the outliers in Figure~\ref{fig:struct_compare}, so we retained them in the sample.

Our initial model was for $P({\rm LMXB})$ as a function of four variables, 
${\rm [Fe/H]}$, $\log_{10} R_{\rm gc}$, $\log_{10} M$, and $\log_{10} \Gamma$. The function is
\begin{equation}
P({\rm LMXB}) = \frac{\exp(\beta_0 + \beta_1 {\rm [Fe/H]}+\beta_2 \log R_{\rm gc} + \beta_3 \log M + \beta_4 \log \Gamma)}
{1+\exp(\beta_0 + \beta_1 {\rm [Fe/H]}+\beta_2 \log R_{\rm gc} + \beta_3 \log M + \beta_4 \log \Gamma)}
\end{equation}
with the coefficients $\beta$ and their standard errors given as model 1 in Table~\ref{tab:logit}.
(The intercept $\beta_0$ is a measure of the sampling proportions in the data and has no physical significance.)
While our sample is defined by the available {\em HST} data and we cannot entirely remove sample bias,
we tried to check for biases by repeating the model-fitting 
for 500 trials where 30 LMXB and 30 non--LMXB clusters were selected at random from
our full sample, and for an additional 500 trials with 41 clusters of each type (i.e., the full LMXB sample with
a randomly-selected comparison sample). 
While there was variation over the trials in the magnitudes of $\beta$, their signs and overall order of
significance did not change.

As expected, $\Gamma$ had
a strong positive effect on $P({\rm LMXB})$ --- increasing $\Gamma$ by a factor of 10, while keeping
all other variables constant, increases the odds of a cluster containing an LMXB by a multiplicative factor of $\exp(\beta_4) = 138$.
Distance from the center of the galaxy $R_{\rm gc}$ also affects $P({\rm LMXB})$ in the way expected from
Figure~\ref{fig:x_compare}: clusters further from the center of the galaxy are less likely to contain an LMXB.
Interestingly, $\beta_1$, the coefficient corresponding to ${\rm [Fe/H]}$, is not significantly
different from zero at the $p<0.05$ level. 
This is consistent with the results of \citetalias{peacock10}, who also found that metallicity was a less important
contributor to $P({\rm LMXB})$ than cluster luminosity or collision rate.
The lack of dependence on ${\rm [Fe/H]}$ could be due to the biases in our sample, or could indicate
that the relationships between ${\rm [Fe/H]}$ and the other variables capture all of the variation due to metallicity.
Also interesting is that
$\beta_3$, the coefficient corresponding to $\log M$, is negative---that is,
for two clusters with the same collision rate $\Gamma$, the higher-mass cluster has a lower chance of containing an LMXB.
The latter finding is consistent with Figure~\ref{fig:scr_mass}, in that the offset in $\Gamma$ between
LMXB and non-LMXB clusters is reduced at higher masses.

\begin{figure}
\plottwo{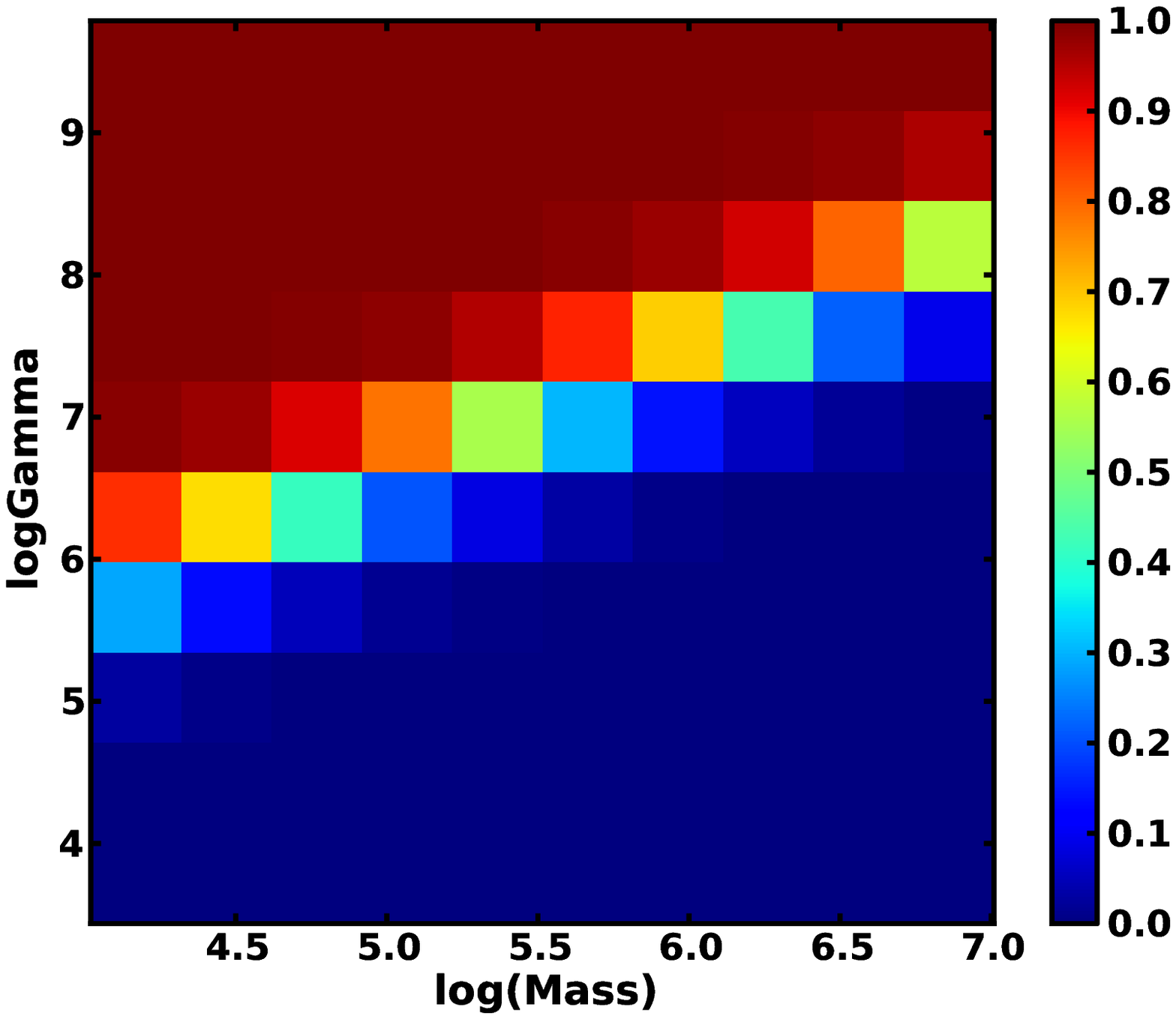}{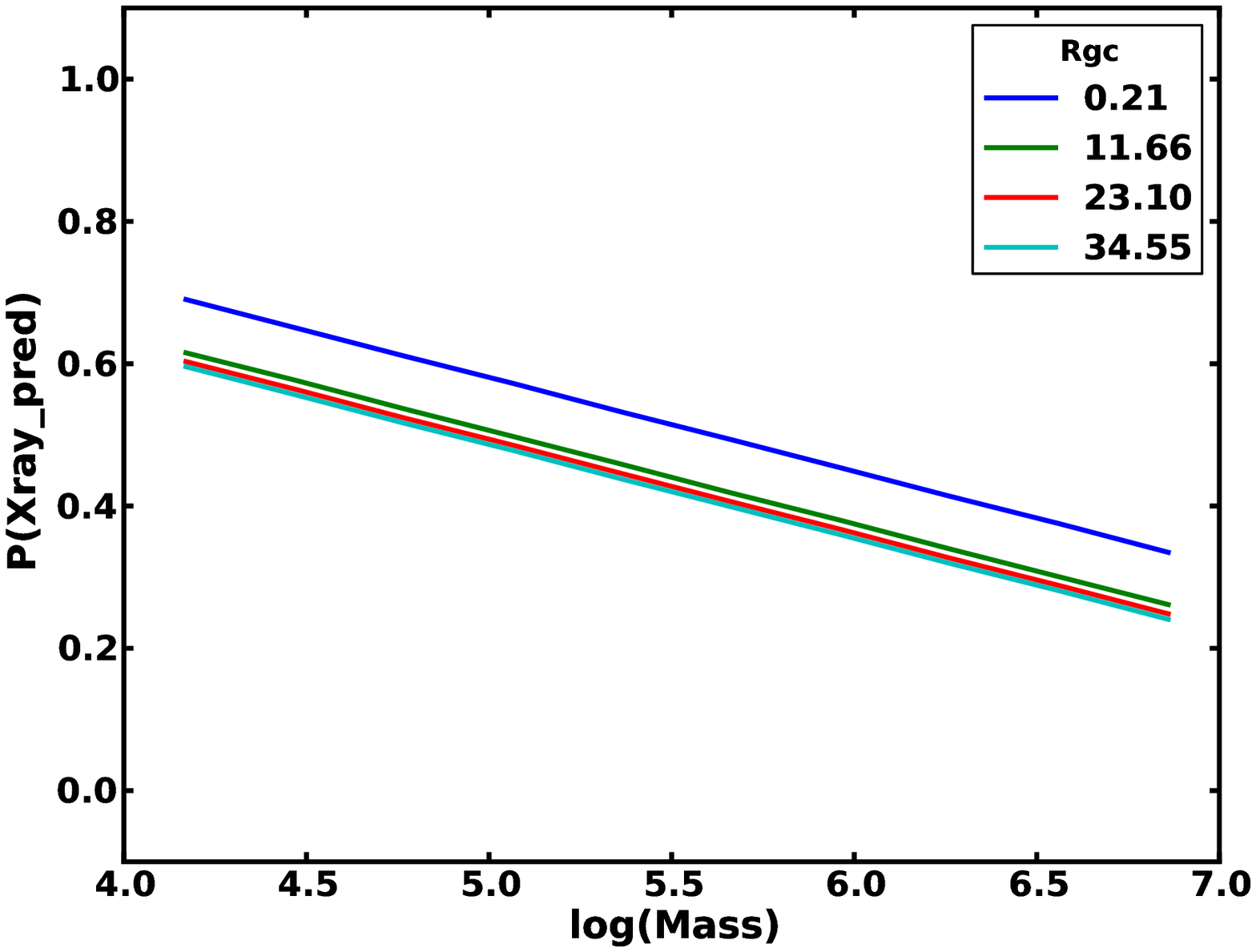}
\caption{
Left: Probability $P_{\rm LMXB}$ predicted from logistic regression model (2) as a function of
$\log_{10} M$, and $\log_{10} \Gamma$, averaging over the range of $R_{\rm gc}$.
Right: $P_{\rm LMXB}(M)$ for different values of $R_{\rm gc}$, averaging over $\Gamma$.
\label{fig:logit}}
\end{figure}

Given the non-significance of a dependence on ${\rm [Fe/H]}$, we re-fit the logistic model including only  
$\log R_{\rm gc}$, $\log M$, and $\log \Gamma$ (model 2 in Table~\ref{tab:logit}). Both the 
Akaike and Bayesian information criteria%
\footnote{
The AIC and BIC are two commonly used criteria for deciding which of a set of models adequately describes
a regression, while giving preference to models with fewer parameters \citep{abraham06}.}
favored the three-parameter fit over the four-parameter one, although Table~\ref{tab:logit} shows that the parameters for the
two types are not greatly different. 
Given the correlation between mass and $\Gamma$ (Figure~\ref{fig:scr_mass}),  it is reasonable to ask whether both
are needed in the logistic regression.  We performed  two-parameter fits with $\log R_{\rm gc}$
and either $\log M$ or $\log \Gamma$, but the information criteria used above favored the three-parameter
fit in both cases.
Returning to the three-parameter fit, we again ran the 500-trial model-fitting exercise with a random selection
of non-LMXB clusters. The distribution of parameter values resulting from the 500 fits
(model 3 in Table~\ref{tab:logit}) matches the 95\% confidence intervals for the full-dataset fit reasonably well.
To summarize the results graphically, we used the three-parameter fit (model 2) results
to predict $P_{\rm LMXB}$ on a grid of values spanning the observed ranges in $\log R_{\rm gc}$, $\log M$, and $\log \Gamma$. 
The left panel of Figure~\ref{fig:logit}  shows $P_{\rm LMXB}$  as a function of $\log M$, and $\log \Gamma$, 
averaging over the $\log R_{\rm gc}$ range, while the right panel demonstrates the relatively small
effect of changes in $\log R_{\rm gc}$ on $P_{\rm LMXB}(M)$. 

What do the regression results mean, physically? The positive dependence on collision rate supports
the idea that dynamical interactions are responsible for LMXB formation. The weak negative dependence
on  galactocentric distance could be a result of the relations between cluster position and both structure
and metallicity.  Having all metallicity dependence be related to cluster position 
would be different from the results of \citet{kim13}, who found that, for elliptical galaxies in Fornax and Virgo, the LMXB
enhancement in metal-rich clusters is not affected by cluster mass, position, or collision rate
[although the collision rate used by \citet{kim13} is $\Gamma_h$, which \citet{maccarone11}
found to be unreliable.] \citet{ivanova12} suggested that metallicity effects on the red giant populations
are the underlying cause of the  LMXB enhancement. 
There are differences between the globular cluster population of M31 studied here and the GCs of cluster elliptical 
galaxies which might explain the lack of metallicity effect on $P_{\rm LMXB}$ in M31.
For example, unlike for most elliptical galaxies, the M31 globular cluster system metallicity distribution
does not appear to be bimodal \citep[][and see Figure~\ref{fig:x_compare}, which however shows only a subsample of M31 globulars]{caldwell11}.
Position-dependent selection effects in the M31 sample could also be biasing our results.  
The available M31 X--ray data restrict our study to GCs with much smaller galactocentric distances than
is typical for studies of ellipticals, meaning that the number of blue metal-poor GCs in our sample is also comparatively lower.
This would make a metallicity dependence of $P_{\rm LMXB}$ in M31 more difficult to observe.

Sample selection biases might also explain the dependence we
find for $P_{\rm LMXB}$ on mass. Higher-mass M31 clusters were more likely to be selected for
targeted {\em HST} observations, and thus have available data regardless of whether or not
they were LMXB hosts. Since we searched for only LMXB-hosting clusters in serendipitous
observations (which would be more likely to include lower-mass clusters), lower-mass clusters
could be overrepresented in our LMXB sample. 
A full examination of the effects of cluster mass
on  $P_{\rm LMXB}$, and the interactions between mass and collision rate, will be possible once
an unbiased sample of M31 clusters with {\em HST} imaging is available; \citet{johnson12}
showed a preview of what will be possible with the full PHAT dataset.

\begin{deluxetable}{lrrr}
\tablewidth{0pt}
\tablecaption{Logistic Regression Coefficients for $P_{\rm LMXB}$ \label{tab:logit}}
\tablehead{
\colhead{variable} & \colhead{Model 1} & \colhead{Model 2} & \colhead{Model 3}}
\startdata
$\beta_1$: ${\rm [Fe/H]}$              &  $-0.32\pm0.55$ & \nodata & \nodata \\
$\beta_2$:  $\log R_{\rm gc}$       & $ -1.34\pm 0.63$   & $-1.25\pm 0.62$& $-1.28 \pm 0.33$\\
$\beta_3$:  $\log M$                       & $ -4.14\pm 1.35$  & $ -3.88\pm 1.25$  & $ -4.08\pm0.77$ \\
$\beta_4$: $\log \Gamma$            & $ 4.93\pm1.27$ & $ 4.64\pm 1.14$ & $ 4.89\pm0.72$ \\
\enddata
\tablecomments{
Model 1 is the four-parameter fit, model 2 the three-parameter fit, and model 3 the
average of $\beta$ values from 500 three-parameter fits on using different random selections
of non-LMXB clusters. 
Uncertainties are standard errors of the coefficients for models 1 and 2, and
standard deviations of the results over 500 trials for model 3.}
\end{deluxetable}

\begin{figure}
\plotone{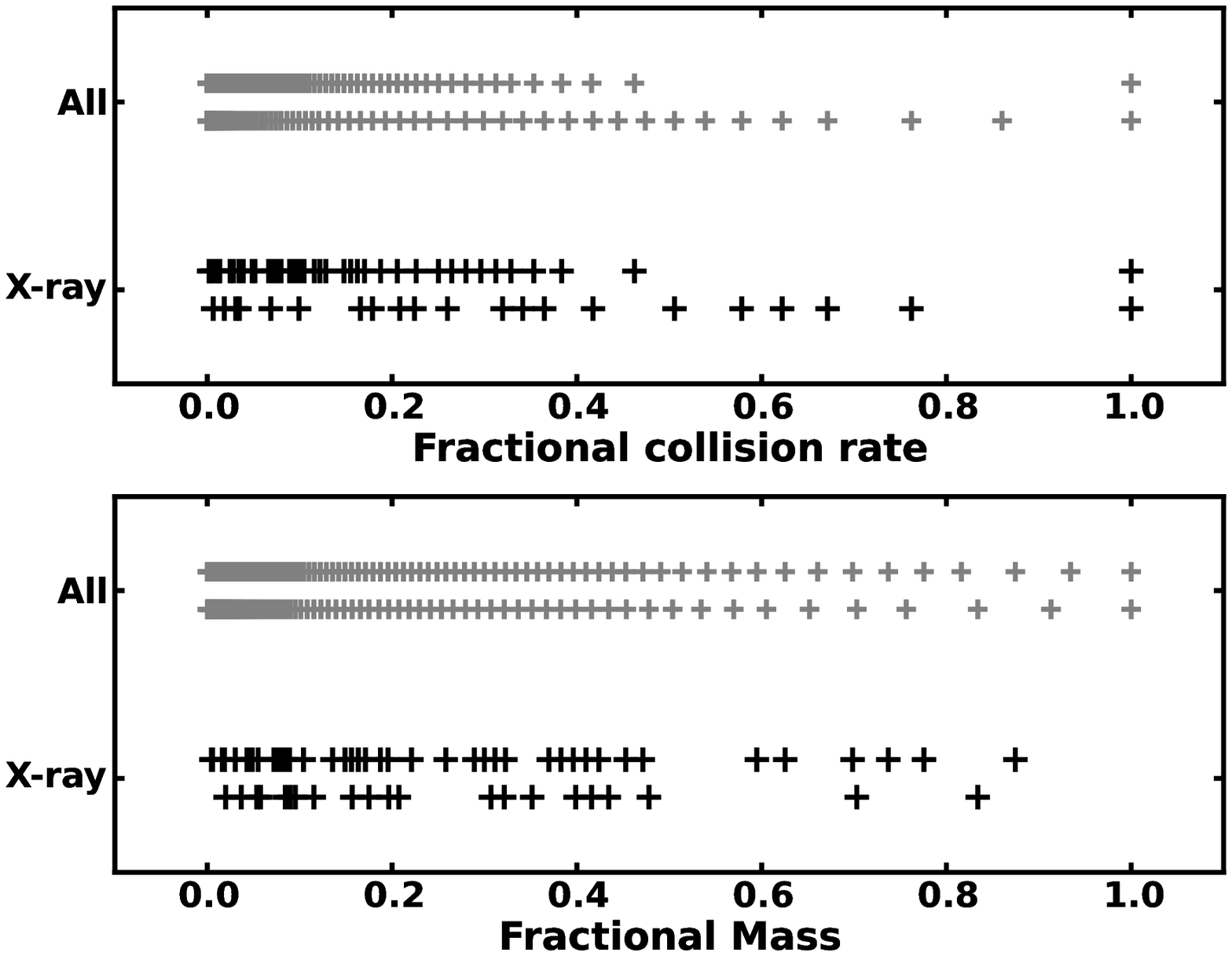}
\caption{
Fractional stellar collision rate and mass distributions for LMXB- and non-LMXB hosting clusters.
The top row of points for each distribution shows the full dataset while the bottom row shows the effect
of omitting the LMXB clusters in the lower half of the X--ray luminosity distribution. 
Cluster B138 has an extreme value of the collision rate (see Section~\ref{sec:modfit}) but a low $L_X$ 
and so affects the top rows in the upper panel.
\label{fig:frac_scr}}
\end{figure}

Finally, we consider the issue of whether the probability of LMXB formation is linearly proportional to collision rate.
The logistic model does not have the correct functional form to address this, so
we instead followed the method of \citet{verbunt87}, also used by \citetalias{peacock10}.  We determined the 
total collision rate $\Gamma_t= \sum_i \Gamma_i$, then sorted the clusters by $\Gamma_i$
and determined the cumulative rate for each cluster  $C_{\Gamma,j} = \sum_0^{<j} \Gamma_i/\Gamma_t$. 
If the $P_{\rm LMXB} \propto \Gamma$ then one would expect that the 
clusters containing (e.g.) 25\% of the total collisions would also contain 25\% of the LMXBs,
so the LMXB clusters should be evenly distributed in $C_{\Gamma,j}$.
This analysis is sensitive to the highest computed collision rates, since they disproportionately affect
$\Gamma_t$, and  as  \citetalias{peacock10} pointed out, can also be problematic if 
the X--ray brightest clusters have more than one LMXB.
Figure~\ref{fig:frac_scr} shows the distribution of fractional collision rate and fractional mass
(computed as  $C_{M,j} = \sum_0^{<j} M_i/M_t$) for both LMXB and non-LMXB clusters. 
We computed the fractional rates using both the full
LMXB sample and the clusters in the upper half of the X--ray luminosity distribution
($L_X\geq 2\times10^{37}$~erg~s$^{-1}$) to more closely match observations in more distant galaxies.
KS tests  reject the hypothesis that any of these distributions matches a uniform one at the $p<0.001$ level,
with the exception of  $C_{\Gamma,j}$ for the X--ray luminous clusters ($p=0.034$).

Results of the fractional rate method are not straightforward to interpret statistically. 
The approximate match of the fractional collision rate distribution for the high-$L_X$
clusters appears to be consistent with the expectation that the LMXB clusters should be
evenly distributed in $C_{\Gamma,j}$, and
we conclude that there is no strong evidence from the fractional rates to support a claim of
non-linearity in $P_{\rm LMXB}(\Gamma)$. The caveat of \citet{maccarone11} that core radius 
measurement errors can easily result in erroneous non-linear $P_{\rm LMXB}(\Gamma)$ relations,
combined with the  possible degeneracies in our fits and biases in our sample, 
would make such an effect difficult to detect in any case. We suggest that a more sophisticated
regression model combined  with a complete sample of M31 clusters would be a better way to approach
the question of linear dependence of  $P_{\rm LMXB}$ on collision rate.

\section{Summary}

A search of the Hubble Legacy Archive yielded 29  M31 GCs 
for which structural parameters had not previously been analyzed in the context
of LMXB association. Combining our fits to the surface brightness profiles
of these clusters with data from the literature, we compiled a sample of 41 LMXB-hosting 
clusters and a comparison sample of 65 non-LMXB clusters.  
The LMXB cluster sample typically has higher metallicity and lower
distance to the center of M31, but this is at least partly due to selection effects.
Assuming a dynamical formation scenario for LMXBs, stellar collision rate can  provide an indication of how likely a cluster is
to host an LMXB.  We find that the LMXB-hosting GCs in M31
have, on average, stellar collision rates a factor of 6 higher  than non-LMXB clusters of the same mass.
A logistic regression analysis indicates that  $P_{\rm LMXB}$ has a weak negative dependence
on galactocentric distance, negligible dependence on metallicity, a strong positive dependence on collision rate, and 
a negative dependence on mass, although the correlation between cluster
mass and collision rate means that LMXB clusters are still more likely to have higher masses.

Although our results are generally in agreement with previous results, they also raise new questions.
The high stellar collision rate  seen in clusters with LMXBs agrees with the dynamical formations scenario for LMXBs,
and the weak effect of $R_{\rm gc}$ on $P_{\rm LMXB}$ is consistent with other work on cluster structure.
The finding that increasing mass decreases  $P_{\rm LMXB}$ is new, however, and it will be
interesting to find out if such a dependence is seen in a complete sample.
A more complete picture of the M31 LMXB-hosting clusters
will be available with the completion of the PHAT survey.  Other future work that may be carried out
with this dataset includes the localization of LMXBs within M31 globulars,  as was done for 
the large M31 cluster G1 by \citet{kong10}. Future high-resolution imaging (for example,
using large ground-based telescopes with adaptive optics)  may even permit the identification of 
optical/near-infrared counterparts to the M31 GC LMXBs. This would shed further light on
the formation pathway of X-ray binaries in these dense stellar systems.

\acknowledgments
We thank the referee for constructive criticism and substantial patience.
We thank M. Gracie for help with HLA searches, T. Desjardins for help
with  synphot, L.C. Johnson for extracting cluster images from PHAT, and E. Cameron
for suggesting we pursue logistic regression.
N. Vulic and C. Heinke provided  helpful comments on the manuscript.
Financial support from an NSERC Discovery Grant and 
an Ontario Early Researcher Award is acknowledged.

{\it Facility:} \facility{HST (WFPC2, ACS)}

\bibliographystyle{apj}

\addtocounter{table}{-3}
\begin{deluxetable}{lllccrrrrrrrr}
\tabletypesize{\scriptsize}
\rotate
\tablecolumns{13}
\tablewidth{0pt}
\tablecaption{Model-fitting results: full version
}
\tablehead{
\colhead{Name} & \colhead{Camera/} &
\colhead{$N_{\rm pts}$} & \colhead{Model} & \colhead{$\chi_{\rm min}^2$}   &
\colhead{$I_{\rm bkg}$} & \colhead{$W_0$} & \colhead{$c$}         &
\colhead{$\mu_0$} & \colhead{$\log\,r_0$} & \colhead{$\log\,r_0$}         \\
\colhead{} & \colhead{Filter}  & \colhead{}   &
\colhead{} & \colhead{} & \colhead{[$L_\odot\,{\rm pc}^{-2}$]} & \colhead{}  &
\colhead{} & \colhead{[mag arcsec$^{-2}$]} & \colhead{[arcsec]}              &
\colhead{[pc]} }
\startdata
\object[SKHV 148]{B086-G148}  & WFC/F435   & $26$       & K66  & $321.95$  & $240.42\pm35.61$  & $7.50^{+0.20}_{-0.20}$  & $1.68^{+0.06}_{-0.06}$  & $14.64^{+0.01}_{-0.01}$  & $-0.709^{+0.009}_{-0.010}$  & $-0.130^{+0.009}_{-0.010}$ \\
          ~~        & ~~      & & W  & $482.40$  & $148.41\pm56.15$  & $7.90^{+0.30}_{-0.30}$  & $3.10^{+0.15}_{-0.22}$  & $14.65^{+0.01}_{-0.01}$  & $-0.698^{+0.012}_{-0.011}$  & $-0.119^{+0.012}_{-0.011}$ \\
          ~~        & ~~     & & PL  & $438.38$  & $161.08\pm10.61$       & ---  & $3.10^{+0.03}_{-0.00}$  & $14.64^{+0.00}_{-0.00}$  & $-0.706^{+0.008}_{-0.000}$  & $-0.127^{+0.008}_{-0.000}$ \\
          ~~        & ~~    & & K62  & $368.87$  & $180.93\pm68.68$       & ---  & $2.08^{+0.82}_{-0.30}$  & $14.64^{+0.01}_{-0.01}$  & $-0.725^{+0.010}_{-0.009}$  & $-0.145^{+0.010}_{-0.009}$ \\
\object[Bol D091]{B091D-D057}  & WFC/F555   & $28$       & K66  & $315.28$  & $93.88\pm4.27$  & $15.90^{+0.10}_{-0.70}$  & $3.56^{+0.02}_{-0.16}$  & $7.91^{+0.34}_{-0.05}$  & $-2.918^{+0.140}_{-0.020}$  & $-2.339^{+0.140}_{-0.020}$ \\
          ~~        & ~~      & & W  & $832.33$  & $31.97\pm8.23$  & $14.60^{+0.50}_{-0.60}$  & $4.08^{+0.13}_{-0.15}$  & $8.32^{+0.24}_{-0.21}$  & $-2.699^{+0.107}_{-0.091}$  & $-2.120^{+0.107}_{-0.091}$ \\
          ~~        & ~~     & & PL  & $1337.80$  & $9.66\pm8.26$       & ---  & $3.10^{+0.03}_{-0.00}$  & $12.77^{+0.01}_{-0.00}$  & $-1.043^{+0.011}_{-0.000}$  & $-0.464^{+0.011}_{-0.000}$ \\
          ~~        & ~~    & & K62  & $1116.00$  & $34.07\pm56.49$       & ---  & $2.24^{+0.82}_{-0.30}$  & $12.73^{+0.06}_{-0.05}$  & $-1.072^{+0.025}_{-0.021}$  & $-0.493^{+0.025}_{-0.021}$ \\
\object[Bol D091]{B091D-D057}  & WFC/F814  & $28$       & K66  & $932.59$  & $125.13\pm25.04$  & $6.70^{+0.50}_{-0.20}$  & $1.44^{+0.15}_{-0.06}$  & $12.90^{+0.01}_{-0.03}$  & $-0.726^{+0.012}_{-0.035}$  & $-0.147^{+0.012}_{-0.035}$ \\
          ~~        & ~~      & & W  & $829.48$  & $105.77\pm15.39$  & $6.40^{+0.10}_{-0.30}$  & $2.00^{+0.06}_{-0.15}$  & $12.91^{+0.01}_{-0.00}$  & $-0.689^{+0.013}_{-0.006}$  & $-0.110^{+0.013}_{-0.006}$ \\
          ~~        & ~~     & & PL  & $869.23$  & $77.47\pm24.84$       & ---  & $3.55^{+0.20}_{-0.10}$  & $12.91^{+0.02}_{-0.01}$  & $-0.662^{+0.039}_{-0.023}$  & $-0.083^{+0.039}_{-0.023}$ \\
          ~~        & ~~    & & K62  & $948.52$  & $115.02\pm20.51$       & ---  & $1.46^{+0.08}_{-0.12}$  & $12.88^{+0.01}_{-0.01}$  & $-0.750^{+0.017}_{-0.011}$  & $-0.171^{+0.017}_{-0.011}$ \\
\object[2MASXi J0042250+405717]{B094-G156}  & WFC/F555   & $26$       & K66   & $61.82$  & $13.75\pm5.94$  & $7.80^{+0.10}_{-0.10}$  & $1.77^{+0.03}_{-0.03}$  & $14.15^{+0.02}_{-0.02}$  & $-0.897^{+0.008}_{-0.008}$  & $-0.318^{+0.008}_{-0.008}$ \\
          ~~        & ~~      & & W  & $170.30$  & $-22.37\pm12.40$  & $7.90^{+0.20}_{-0.10}$  & $3.10^{+0.11}_{-0.07}$  & $14.19^{+0.01}_{-0.02}$  & $-0.870^{+0.006}_{-0.012}$  & $-0.290^{+0.006}_{-0.012}$ \\
          ~~        & ~~     & & PL  & $190.94$  & $-20.31\pm1.37$       & ---  & $3.10^{+0.01}_{-0.00}$  & $14.19^{+0.00}_{-0.00}$  & $-0.876^{+0.003}_{-0.000}$  & $-0.297^{+0.003}_{-0.000}$ \\
          ~~        & ~~    & & K62  & $118.07$  & $-16.86\pm13.79$       & ---  & $2.32^{+0.40}_{-0.18}$  & $14.17^{+0.01}_{-0.01}$  & $-0.904^{+0.005}_{-0.006}$  & $-0.324^{+0.005}_{-0.006}$ \\
\object[NBol 002]{B096-G158}  & WFC/F814  & $26$       & K66   & $44.45$  & $780.22\pm9.26$  & $6.60^{+0.20}_{-0.10}$  & $1.41^{+0.06}_{-0.03}$  & $13.61^{+0.01}_{-0.02}$  & $-0.724^{+0.007}_{-0.015}$  & $-0.145^{+0.007}_{-0.015}$ \\
          ~~        & ~~      & & W   & $58.69$  & $752.77\pm14.28$  & $6.60^{+0.20}_{-0.20}$  & $2.11^{+0.13}_{-0.12}$  & $13.63^{+0.01}_{-0.02}$  & $-0.695^{+0.013}_{-0.013}$  & $-0.115^{+0.013}_{-0.013}$ \\
          ~~        & ~~     & & PL   & $68.31$  & $736.01\pm22.23$       & ---  & $3.50^{+0.10}_{-0.15}$  & $13.64^{+0.02}_{-0.03}$  & $-0.663^{+0.023}_{-0.039}$  & $-0.084^{+0.023}_{-0.039}$ \\
          ~~        & ~~    & & K62   & $43.89$  & $776.74\pm10.95$       & ---  & $1.38^{+0.06}_{-0.06}$  & $13.60^{+0.01}_{-0.01}$  & $-0.741^{+0.011}_{-0.010}$  & $-0.162^{+0.011}_{-0.010}$ \\
\object[NBol 001]{B107-G169}  & WFC/F475   & $26$       & K66  & $323.94$  & $622.06\pm19.11$  & $7.90^{+0.20}_{-0.20}$  & $1.80^{+0.06}_{-0.06}$  & $13.93^{+0.06}_{-0.06}$  & $-0.932^{+0.026}_{-0.027}$  & $-0.352^{+0.026}_{-0.027}$ \\
          ~~        & ~~      & & W  & $540.16$  & $555.12\pm47.02$  & $8.00^{+0.40}_{-0.30}$  & $3.15^{+0.14}_{-0.20}$  & $14.01^{+0.06}_{-0.09}$  & $-0.891^{+0.029}_{-0.039}$  & $-0.312^{+0.029}_{-0.039}$ \\
          ~~        & ~~     & & PL  & $560.10$  & $567.59\pm7.05$       & ---  & $3.10^{+0.03}_{-0.00}$  & $14.06^{+0.02}_{-0.00}$  & $-0.881^{+0.012}_{-0.000}$  & $-0.302^{+0.012}_{-0.000}$ \\
          ~~        & ~~    & & K62  & $480.94$  & $563.57\pm49.11$       & ---  & $2.50^{+0.70}_{-0.56}$  & $14.00^{+0.05}_{-0.02}$  & $-0.919^{+0.025}_{-0.010}$  & $-0.340^{+0.025}_{-0.010}$ \\
\object[NBol 001]{B107-G169}  & WFC/F814  & $26$       & K66  & $124.62$  & $879.24\pm44.26$  & $7.90^{+0.30}_{-0.20}$  & $1.80^{+0.09}_{-0.06}$  & $12.88^{+0.02}_{-0.04}$  & $-0.879^{+0.014}_{-0.021}$  & $-0.299^{+0.014}_{-0.021}$ \\
          ~~        & ~~      & & W  & $200.71$  & $773.48\pm57.01$  & $8.30^{+0.30}_{-0.30}$  & $3.27^{+0.05}_{-0.12}$  & $12.90^{+0.02}_{-0.03}$  & $-0.864^{+0.016}_{-0.016}$  & $-0.284^{+0.016}_{-0.016}$ \\
          ~~        & ~~     & & PL  & $236.11$  & $832.59\pm6.28$       & ---  & $3.10^{+0.02}_{-0.00}$  & $12.94^{+0.01}_{-0.00}$  & $-0.848^{+0.007}_{-0.000}$  & $-0.269^{+0.007}_{-0.000}$ \\
          ~~        & ~~    & & K62  & $180.30$  & $804.10\pm36.45$       & ---  & $3.16^{+0.04}_{-0.90}$  & $12.90^{+0.01}_{-0.00}$  & $-0.883^{+0.009}_{-0.001}$  & $-0.304^{+0.009}_{-0.001}$ \\
\object[NBol 001]{B107-G169}  & WFPC/F606  & $29$       & K66   & $86.72$  & $305.29\pm16.43$  & $7.30^{+0.20}_{-0.30}$  & $1.62^{+0.06}_{-0.09}$  & $13.01^{+0.04}_{-0.03}$  & $-0.747^{+0.022}_{-0.015}$  & $-0.167^{+0.022}_{-0.015}$ \\
          ~~        & ~~      & & W  & $112.56$  & $286.05\pm24.72$  & $7.10^{+0.30}_{-0.30}$  & $2.46^{+0.25}_{-0.22}$  & $13.06^{+0.03}_{-0.03}$  & $-0.699^{+0.019}_{-0.019}$  & $-0.120^{+0.019}_{-0.019}$ \\
          ~~        & ~~     & & PL  & $124.49$  & $267.97\pm25.78$       & ---  & $3.30^{+0.15}_{-0.10}$  & $13.06^{+0.04}_{-0.03}$  & $-0.691^{+0.041}_{-0.031}$  & $-0.112^{+0.041}_{-0.031}$ \\
          ~~        & ~~    & & K62   & $94.07$  & $305.31\pm16.91$       & ---  & $1.62^{+0.12}_{-0.10}$  & $13.02^{+0.02}_{-0.02}$  & $-0.748^{+0.013}_{-0.012}$  & $-0.168^{+0.013}_{-0.012}$ \\
\object[SKHV 178]{B116-G178}  & WFC/F555   & $28$       & K66   & $78.26$  & $23.79\pm11.94$  & $7.60^{+0.10}_{-0.20}$  & $1.71^{+0.03}_{-0.06}$  & $13.61^{+0.04}_{-0.02}$  & $-0.899^{+0.020}_{-0.010}$  & $-0.319^{+0.020}_{-0.010}$ \\
          ~~        & ~~      & & W  & $192.64$  & $-12.77\pm20.09$  & $7.50^{+0.20}_{-0.20}$  & $2.79^{+0.16}_{-0.17}$  & $13.67^{+0.03}_{-0.02}$  & $-0.853^{+0.015}_{-0.015}$  & $-0.273^{+0.015}_{-0.015}$ \\
          ~~        & ~~     & & PL  & $196.26$  & $-34.25\pm12.61$       & ---  & $3.13^{+0.06}_{-0.03}$  & $13.64^{+0.03}_{-0.01}$  & $-0.875^{+0.023}_{-0.012}$  & $-0.296^{+0.023}_{-0.012}$ \\
          ~~        & ~~    & & K62  & $121.28$  & $12.02\pm16.72$       & ---  & $1.82^{+0.12}_{-0.08}$  & $13.64^{+0.01}_{-0.02}$  & $-0.892^{+0.008}_{-0.009}$  & $-0.312^{+0.008}_{-0.009}$ \\
\object[SKHV 178]{B116-G178}  & WFC/F814  & $28$       & K66   & $83.04$  & $30.58\pm3.98$  & $7.50^{+0.10}_{-0.10}$  & $1.68^{+0.03}_{-0.03}$  & $12.63^{+0.03}_{-0.04}$  & $-0.867^{+0.014}_{-0.015}$  & $-0.287^{+0.014}_{-0.015}$ \\
          ~~        & ~~      & & W  & $210.09$  & $1.65\pm14.46$  & $7.20^{+0.20}_{-0.20}$  & $2.54^{+0.17}_{-0.16}$  & $12.70^{+0.04}_{-0.04}$  & $-0.813^{+0.021}_{-0.022}$  & $-0.234^{+0.021}_{-0.022}$ \\
          ~~        & ~~     & & PL  & $266.63$  & $-14.01\pm17.57$       & ---  & $3.30^{+0.10}_{-0.10}$  & $12.71^{+0.06}_{-0.06}$  & $-0.799^{+0.037}_{-0.042}$  & $-0.220^{+0.037}_{-0.042}$ \\
          ~~        & ~~    & & K62  & $119.26$  & $22.93\pm6.75$       & ---  & $1.72^{+0.04}_{-0.06}$  & $12.66^{+0.03}_{-0.02}$  & $-0.862^{+0.012}_{-0.008}$  & $-0.282^{+0.012}_{-0.008}$ \\
\object[SKHV 176]{B117-G176}  & WFPC/F336   & $29$       & K66  & $105.81$  & $303.29\pm117.83$  & $9.10^{+0.30}_{-0.30}$  & $2.14^{+0.07}_{-0.08}$  & $13.85^{+0.04}_{-0.05}$  & $-0.910^{+0.018}_{-0.018}$  & $-0.330^{+0.018}_{-0.018}$ \\
          ~~        & ~~      & & W  & $123.52$  & $124.59\pm126.76$  & $9.80^{+0.40}_{-0.30}$  & $3.32^{+0.01}_{-0.00}$  & $13.88^{+0.04}_{-0.03}$  & $-0.894^{+0.016}_{-0.016}$  & $-0.315^{+0.016}_{-0.016}$ \\
          ~~        & ~~     & & PL  & $328.42$  & $540.06\pm17.31$       & ---  & $3.10^{+0.03}_{-0.00}$  & $14.29^{+0.00}_{-0.00}$  & $-0.755^{+0.007}_{-0.000}$  & $-0.176^{+0.007}_{-0.000}$ \\
          ~~        & ~~    & & K62  & $265.26$  & $498.84\pm23.36$       & ---  & $3.20^{+0.00}_{-0.68}$  & $14.21^{+0.02}_{-0.01}$  & $-0.801^{+0.007}_{-0.003}$  & $-0.221^{+0.007}_{-0.003}$ \\
\object[SKHV 187]{B128-G187}  & WFPC/F555   & $22$       & K66    & $5.59$  & $102.16\pm3.16$  & $7.70^{+0.10}_{-0.10}$  & $1.74^{+0.03}_{-0.03}$  & $14.70^{+0.03}_{-0.03}$  & $-0.997^{+0.012}_{-0.012}$  & $-0.418^{+0.012}_{-0.012}$ \\
          ~~        & ~~      & & W   & $16.57$  & $84.12\pm8.67$  & $7.80^{+0.20}_{-0.20}$  & $3.03^{+0.13}_{-0.16}$  & $14.78^{+0.04}_{-0.04}$  & $-0.953^{+0.018}_{-0.018}$  & $-0.374^{+0.018}_{-0.018}$ \\
          ~~        & ~~     & & PL   & $19.24$  & $83.11\pm1.58$       & ---  & $3.10^{+0.02}_{-0.00}$  & $14.78^{+0.02}_{-0.00}$  & $-0.964^{+0.009}_{-0.000}$  & $-0.385^{+0.009}_{-0.000}$ \\
          ~~        & ~~    & & K62   & $14.64$  & $88.99\pm8.94$       & ---  & $2.16^{+0.32}_{-0.16}$  & $14.75^{+0.02}_{-0.02}$  & $-0.994^{+0.009}_{-0.011}$  & $-0.414^{+0.009}_{-0.011}$ \\
\object[SKHV 187]{B128-G187}  & WFPC/F814  & $22$       & K66   & $20.37$  & $167.96\pm8.51$  & $7.20^{+0.20}_{-0.20}$  & $1.59^{+0.06}_{-0.06}$  & $13.94^{+0.03}_{-0.03}$  & $-0.910^{+0.016}_{-0.016}$  & $-0.331^{+0.016}_{-0.016}$ \\
          ~~        & ~~      & & W   & $16.83$  & $149.76\pm10.54$  & $7.30^{+0.20}_{-0.20}$  & $2.62^{+0.17}_{-0.16}$  & $13.98^{+0.02}_{-0.02}$  & $-0.877^{+0.013}_{-0.013}$  & $-0.298^{+0.013}_{-0.013}$ \\
          ~~        & ~~     & & PL   & $17.11$  & $144.65\pm10.14$       & ---  & $3.25^{+0.05}_{-0.09}$  & $13.99^{+0.02}_{-0.04}$  & $-0.871^{+0.016}_{-0.031}$  & $-0.292^{+0.016}_{-0.031}$ \\
          ~~        & ~~    & & K62   & $17.25$  & $158.23\pm10.04$       & ---  & $1.74^{+0.10}_{-0.12}$  & $13.93^{+0.02}_{-0.01}$  & $-0.928^{+0.012}_{-0.008}$  & $-0.349^{+0.012}_{-0.008}$ \\
\object[SKHV 192]{B135-G192}  & WFC/F555   & $28$       & K66  & $146.22$  & $13.12\pm2.06$  & $15.40^{+0.40}_{-0.60}$  & $3.44^{+0.09}_{-0.13}$  & $9.31^{+0.28}_{-0.20}$  & $-2.689^{+0.117}_{-0.082}$  & $-2.110^{+0.117}_{-0.082}$ \\
          ~~        & ~~      & & W  & $373.52$  & $-21.15\pm4.87$  & $15.00^{+0.50}_{-0.70}$  & $4.18^{+0.13}_{-0.18}$  & $9.37^{+0.29}_{-0.22}$  & $-2.625^{+0.125}_{-0.094}$  & $-2.046^{+0.125}_{-0.094}$ \\
          ~~        & ~~     & & PL  & $705.10$  & $-19.63\pm3.31$       & ---  & $3.10^{+0.03}_{-0.00}$  & $13.87^{+0.02}_{-0.00}$  & $-0.923^{+0.015}_{-0.000}$  & $-0.343^{+0.015}_{-0.000}$ \\
          ~~        & ~~    & & K62  & $583.03$  & $-29.46\pm18.10$       & ---  & $3.14^{+0.06}_{-0.88}$  & $13.79^{+0.05}_{-0.01}$  & $-0.973^{+0.022}_{-0.003}$  & $-0.394^{+0.022}_{-0.003}$ \\
\object[SKHV 192]{B135-G192}  & WFC/F814  & $28$       & K66  & $293.74$  & $10.87\pm8.25$  & $8.40^{+0.20}_{-0.20}$  & $1.95^{+0.06}_{-0.06}$  & $12.79^{+0.06}_{-0.07}$  & $-1.023^{+0.028}_{-0.031}$  & $-0.443^{+0.028}_{-0.031}$ \\
          ~~        & ~~      & & W  & $529.65$  & $-24.51\pm19.81$  & $8.30^{+0.30}_{-0.30}$  & $3.27^{+0.05}_{-0.12}$  & $12.85^{+0.07}_{-0.06}$  & $-0.982^{+0.033}_{-0.030}$  & $-0.403^{+0.033}_{-0.030}$ \\
          ~~        & ~~     & & PL  & $672.61$  & $-9.39\pm2.26$       & ---  & $3.10^{+0.02}_{-0.00}$  & $12.97^{+0.01}_{-0.00}$  & $-0.938^{+0.008}_{-0.000}$  & $-0.359^{+0.008}_{-0.000}$ \\
          ~~        & ~~    & & K62  & $546.33$  & $-18.92\pm19.38$       & ---  & $3.20^{+0.00}_{-0.94}$  & $12.90^{+0.04}_{-0.00}$  & $-0.985^{+0.020}_{-0.000}$  & $-0.406^{+0.020}_{-0.000}$ \\
\object[C401141021]{B138}  & WFC/F475   & $26$       & K66   & $43.68$  & $1747.99\pm3.12$  & $10.40^{+0.30}_{-0.20}$  & $2.43^{+0.06}_{-0.04}$  & $10.30^{+0.14}_{-0.19}$  & $-2.291^{+0.054}_{-0.077}$  & $-1.712^{+0.054}_{-0.077}$ \\
          ~~        & ~~      & & W   & $31.95$  & $1727.34\pm1.05$  & $13.00^{+0.60}_{-0.70}$  & $3.70^{+0.13}_{-0.13}$  & $8.59^{+0.24}_{-0.22}$  & $-2.930^{+0.118}_{-0.103}$  & $-2.351^{+0.118}_{-0.103}$ \\
          ~~        & ~~     & & PL  & $130.43$  & $1686.46\pm17.57$       & ---  & $3.14^{+0.11}_{-0.04}$  & $12.92^{+0.15}_{-0.06}$  & $-1.402^{+0.065}_{-0.026}$  & $-0.823^{+0.065}_{-0.026}$ \\
          ~~        & ~~    & & K62  & $101.36$  & $1723.27\pm17.47$       & ---  & $1.96^{+0.20}_{-0.12}$  & $12.89^{+0.06}_{-0.08}$  & $-1.435^{+0.020}_{-0.026}$  & $-0.855^{+0.020}_{-0.026}$ \\
\object[C401540596]{B144}  & WFC/F475   & $26$       & K66   & $30.55$  & $937.70\pm4.41$  & $9.30^{+0.10}_{-0.20}$  & $2.19^{+0.02}_{-0.05}$  & $13.76^{+0.12}_{-0.06}$  & $-1.489^{+0.043}_{-0.022}$  & $-0.910^{+0.043}_{-0.022}$ \\
          ~~        & ~~      & & W   & $39.94$  & $943.82\pm3.67$  & $12.10^{+0.20}_{-0.30}$  & $3.54^{+0.03}_{-0.05}$  & $11.13^{+0.19}_{-0.12}$  & $-2.393^{+0.075}_{-0.047}$  & $-1.814^{+0.075}_{-0.047}$ \\
          ~~        & ~~     & & PL  & $684.46$  & $959.09\pm2.72$       & ---  & $3.10^{+0.02}_{-0.00}$  & $14.40^{+0.01}_{-0.00}$  & $-1.229^{+0.007}_{-0.000}$  & $-0.650^{+0.007}_{-0.000}$ \\
          ~~        & ~~    & & K62  & $516.82$  & $950.50\pm4.76$       & ---  & $3.18^{+0.02}_{-0.48}$  & $14.33^{+0.01}_{-0.00}$  & $-1.276^{+0.004}_{-0.000}$  & $-0.697^{+0.004}_{-0.000}$ \\
\object[C401540596]{B144}  & WFC/F814  & $26$       & K66   & $18.01$  & $1778.37\pm12.37$  & $8.70^{+0.20}_{-0.20}$  & $2.04^{+0.05}_{-0.06}$  & $12.61^{+0.08}_{-0.09}$  & $-1.294^{+0.032}_{-0.035}$  & $-0.715^{+0.032}_{-0.035}$ \\
          ~~        & ~~      & & W   & $22.03$  & $1773.70\pm53.65$  & $15.90^{+0.10}_{-7.20}$  & $4.41^{+0.03}_{-1.09}$  & $7.94^{+4.83}_{-0.04}$  & $-3.053^{+1.832}_{-0.018}$  & $-2.474^{+1.832}_{-0.018}$ \\
          ~~        & ~~     & & PL   & $87.22$  & $1785.87\pm2.72$       & ---  & $3.10^{+0.02}_{-0.00}$  & $13.00^{+0.02}_{-0.00}$  & $-1.136^{+0.010}_{-0.000}$  & $-0.556^{+0.010}_{-0.000}$ \\
          ~~        & ~~    & & K62   & $60.58$  & $1772.18\pm5.21$       & ---  & $3.20^{+0.00}_{-0.44}$  & $12.92^{+0.01}_{-0.01}$  & $-1.183^{+0.004}_{-0.001}$  & $-0.604^{+0.004}_{-0.001}$ \\
\object[C401840589]{B146}   & WFC/F475   & $26$       & K66  & $388.93$  & $614.11\pm28.08$  & $9.10^{+0.50}_{-0.50}$  & $2.14^{+0.12}_{-0.13}$  & $15.48^{+0.06}_{-0.05}$  & $-1.063^{+0.027}_{-0.024}$  & $-0.483^{+0.027}_{-0.024}$ \\
          ~~        & ~~      & & W  & $444.59$  & $585.97\pm31.63$  & $9.80^{+0.70}_{-0.50}$  & $3.32^{+0.03}_{-0.00}$  & $15.50^{+0.04}_{-0.05}$  & $-1.050^{+0.022}_{-0.025}$  & $-0.471^{+0.022}_{-0.025}$ \\
          ~~        & ~~     & & PL  & $764.80$  & $648.12\pm2.43$       & ---  & $3.10^{+0.04}_{-0.00}$  & $15.68^{+0.02}_{-0.00}$  & $-0.969^{+0.014}_{-0.000}$  & $-0.390^{+0.014}_{-0.000}$ \\
          ~~        & ~~    & & K62  & $652.06$  & $640.81\pm7.71$       & ---  & $3.18^{+0.02}_{-0.92}$  & $15.65^{+0.02}_{-0.01}$  & $-0.998^{+0.008}_{-0.003}$  & $-0.419^{+0.008}_{-0.003}$ \\
\object[C401840589]{B146}   & WFC/F814  & $26$       & K66   & $86.08$  & $1139.27\pm29.34$  & $8.50^{+0.40}_{-0.40}$  & $1.98^{+0.11}_{-0.12}$  & $14.21^{+0.05}_{-0.04}$  & $-0.946^{+0.023}_{-0.021}$  & $-0.367^{+0.023}_{-0.021}$ \\
          ~~        & ~~      & & W   & $99.56$  & $1097.01\pm42.73$  & $9.20^{+0.60}_{-0.60}$  & $3.33^{+0.00}_{-0.01}$  & $14.22^{+0.05}_{-0.04}$  & $-0.938^{+0.028}_{-0.024}$  & $-0.358^{+0.028}_{-0.024}$ \\
          ~~        & ~~     & & PL  & $137.76$  & $1158.97\pm3.54$       & ---  & $3.10^{+0.03}_{-0.00}$  & $14.34^{+0.01}_{-0.00}$  & $-0.881^{+0.010}_{-0.000}$  & $-0.302^{+0.010}_{-0.000}$ \\
          ~~        & ~~    & & K62  & $116.40$  & $1149.56\pm11.18$       & ---  & $3.16^{+0.04}_{-0.90}$  & $14.30^{+0.01}_{-0.00}$  & $-0.917^{+0.007}_{-0.001}$  & $-0.337^{+0.007}_{-0.001}$ \\
\object[SKHV 200]{B148-G200}  & WFC/F475   & $26$       & K66  & $276.28$  & $1042.50\pm28.24$  & $7.70^{+0.20}_{-0.30}$  & $1.74^{+0.06}_{-0.09}$  & $14.19^{+0.04}_{-0.03}$  & $-0.928^{+0.024}_{-0.016}$  & $-0.348^{+0.024}_{-0.016}$ \\
          ~~        & ~~      & & W  & $414.60$  & $983.86\pm40.36$  & $7.90^{+0.30}_{-0.30}$  & $3.10^{+0.15}_{-0.22}$  & $14.22^{+0.03}_{-0.03}$  & $-0.903^{+0.019}_{-0.019}$  & $-0.323^{+0.019}_{-0.019}$ \\
          ~~        & ~~     & & PL  & $406.04$  & $991.00\pm8.72$       & ---  & $3.10^{+0.04}_{-0.00}$  & $14.23^{+0.02}_{-0.00}$  & $-0.909^{+0.016}_{-0.000}$  & $-0.330^{+0.016}_{-0.000}$ \\
          ~~        & ~~    & & K62  & $356.25$  & $1010.96\pm56.85$       & ---  & $2.08^{+1.12}_{-0.30}$  & $14.22^{+0.03}_{-0.03}$  & $-0.928^{+0.017}_{-0.017}$  & $-0.349^{+0.017}_{-0.017}$ \\
\object[SKHV 200]{B148-G200}  & WFC/F814  & $26$       & K66  & $127.88$  & $1633.40\pm41.44$  & $7.70^{+0.30}_{-0.20}$  & $1.74^{+0.09}_{-0.06}$  & $12.91^{+0.02}_{-0.03}$  & $-0.924^{+0.012}_{-0.018}$  & $-0.345^{+0.012}_{-0.018}$ \\
          ~~        & ~~      & & W  & $178.78$  & $1552.24\pm64.11$  & $8.00^{+0.40}_{-0.30}$  & $3.15^{+0.14}_{-0.20}$  & $12.93^{+0.02}_{-0.03}$  & $-0.906^{+0.014}_{-0.020}$  & $-0.327^{+0.014}_{-0.020}$ \\
          ~~        & ~~     & & PL  & $178.97$  & $1574.59\pm8.25$       & ---  & $3.10^{+0.03}_{-0.00}$  & $12.93^{+0.01}_{-0.00}$  & $-0.910^{+0.011}_{-0.000}$  & $-0.331^{+0.011}_{-0.000}$ \\
          ~~        & ~~    & & K62  & $151.80$  & $1559.96\pm60.31$       & ---  & $2.54^{+0.66}_{-0.60}$  & $12.91^{+0.02}_{-0.01}$  & $-0.937^{+0.013}_{-0.004}$  & $-0.358^{+0.013}_{-0.004}$ \\
\object[SKHV 203]{B150-G203}  & WFPC/F555   & $24$       & K66   & $20.89$  & $315.88\pm4.72$  & $6.60^{+0.20}_{-0.20}$  & $1.41^{+0.06}_{-0.05}$  & $15.12^{+0.03}_{-0.03}$  & $-0.674^{+0.017}_{-0.017}$  & $-0.095^{+0.017}_{-0.017}$ \\
          ~~        & ~~      & & W   & $14.94$  & $307.48\pm5.63$  & $6.50^{+0.20}_{-0.20}$  & $2.05^{+0.12}_{-0.11}$  & $15.17^{+0.02}_{-0.02}$  & $-0.633^{+0.014}_{-0.014}$  & $-0.054^{+0.014}_{-0.014}$ \\
          ~~        & ~~     & & PL   & $14.86$  & $301.08\pm6.49$       & ---  & $3.55^{+0.10}_{-0.10}$  & $15.19^{+0.03}_{-0.03}$  & $-0.594^{+0.026}_{-0.026}$  & $-0.015^{+0.026}_{-0.026}$ \\
          ~~        & ~~    & & K62   & $19.56$  & $314.63\pm4.40$       & ---  & $1.38^{+0.06}_{-0.06}$  & $15.11^{+0.02}_{-0.02}$  & $-0.692^{+0.013}_{-0.011}$  & $-0.112^{+0.013}_{-0.011}$ \\
\object[SKHV 203]{B150-G203}  & WFPC/F814  & $24$       & K66   & $24.39$  & $379.63\pm9.21$  & $6.60^{+0.30}_{-0.30}$  & $1.41^{+0.09}_{-0.08}$  & $14.31^{+0.03}_{-0.04}$  & $-0.642^{+0.021}_{-0.022}$  & $-0.063^{+0.021}_{-0.022}$ \\
          ~~        & ~~      & & W   & $18.39$  & $367.25\pm10.96$  & $6.60^{+0.30}_{-0.30}$  & $2.11^{+0.20}_{-0.17}$  & $14.34^{+0.03}_{-0.03}$  & $-0.609^{+0.019}_{-0.019}$  & $-0.030^{+0.019}_{-0.019}$ \\
          ~~        & ~~     & & PL   & $17.25$  & $359.39\pm10.32$       & ---  & $3.50^{+0.15}_{-0.10}$  & $14.35^{+0.03}_{-0.03}$  & $-0.577^{+0.035}_{-0.026}$  & $0.002^{+0.035}_{-0.026}$ \\
          ~~        & ~~    & & K62   & $23.43$  & $378.11\pm9.57$       & ---  & $1.38^{+0.12}_{-0.08}$  & $14.30^{+0.02}_{-0.03}$  & $-0.659^{+0.015}_{-0.018}$  & $-0.080^{+0.015}_{-0.018}$ \\
\object[Bol 153]{B153}  & WFC/F475   & $26$       & K66   & $81.42$  & $377.07\pm0.14$  & $16.00^{+0.00}_{-0.10}$  & $3.58^{+0.00}_{-0.02}$  & $8.91^{+0.05}_{-0.00}$  & $-3.102^{+0.021}_{-0.000}$  & $-2.523^{+0.021}_{-0.000}$ \\
          ~~        & ~~      & & W  & $117.12$  & $351.74\pm2.29$  & $14.00^{+0.50}_{-0.40}$  & $3.93^{+0.12}_{-0.09}$  & $9.65^{+0.15}_{-0.20}$  & $-2.736^{+0.070}_{-0.088}$  & $-2.157^{+0.070}_{-0.088}$ \\
          ~~        & ~~     & & PL  & $1047.20$  & $338.41\pm3.65$       & ---  & $3.10^{+0.03}_{-0.00}$  & $14.08^{+0.03}_{-0.00}$  & $-1.136^{+0.013}_{-0.000}$  & $-0.556^{+0.013}_{-0.000}$ \\
          ~~        & ~~    & & K62  & $849.96$  & $328.51\pm19.13$       & ---  & $3.10^{+0.10}_{-0.84}$  & $14.00^{+0.04}_{-0.00}$  & $-1.184^{+0.016}_{-0.000}$  & $-0.604^{+0.016}_{-0.000}$ \\
\object[Bol 153]{B153}  & WFC/F814  & $26$       & K66   & $99.66$  & $621.34\pm13.12$  & $8.00^{+0.20}_{-0.20}$  & $1.83^{+0.06}_{-0.06}$  & $12.55^{+0.08}_{-0.08}$  & $-1.113^{+0.027}_{-0.028}$  & $-0.534^{+0.027}_{-0.028}$ \\
          ~~        & ~~      & & W  & $188.86$  & $569.58\pm26.89$  & $8.00^{+0.20}_{-0.30}$  & $3.15^{+0.09}_{-0.20}$  & $12.66^{+0.07}_{-0.05}$  & $-1.062^{+0.028}_{-0.020}$  & $-0.483^{+0.028}_{-0.020}$ \\
          ~~        & ~~     & & PL  & $213.01$  & $577.49\pm6.37$       & ---  & $3.10^{+0.03}_{-0.00}$  & $12.72^{+0.01}_{-0.00}$  & $-1.054^{+0.010}_{-0.000}$  & $-0.474^{+0.010}_{-0.000}$ \\
          ~~        & ~~    & & K62  & $184.83$  & $587.69\pm39.98$       & ---  & $2.30^{+0.90}_{-0.36}$  & $12.66^{+0.06}_{-0.03}$  & $-1.088^{+0.021}_{-0.014}$  & $-0.509^{+0.021}_{-0.014}$ \\
\object[Bol 159]{B159}  & WFC/F435   & $27$       & K66  & $248.38$  & $-81.05\pm28.79$  & $8.70^{+0.50}_{-0.40}$  & $2.04^{+0.13}_{-0.11}$  & $15.69^{+0.07}_{-0.07}$  & $-0.950^{+0.030}_{-0.031}$  & $-0.371^{+0.030}_{-0.031}$ \\
          ~~        & ~~      & & W  & $301.75$  & $-121.89\pm32.62$  & $9.50^{+0.50}_{-0.60}$  & $3.32^{+0.01}_{-0.00}$  & $15.71^{+0.07}_{-0.05}$  & $-0.940^{+0.035}_{-0.024}$  & $-0.361^{+0.035}_{-0.024}$ \\
          ~~        & ~~     & & PL  & $455.78$  & $-59.75\pm2.64$       & ---  & $3.10^{+0.03}_{-0.00}$  & $15.94^{+0.01}_{-0.00}$  & $-0.848^{+0.010}_{-0.000}$  & $-0.268^{+0.010}_{-0.000}$ \\
          ~~        & ~~    & & K62  & $385.18$  & $-67.80\pm9.65$       & ---  & $3.20^{+0.00}_{-0.94}$  & $15.89^{+0.02}_{-0.01}$  & $-0.884^{+0.007}_{-0.003}$  & $-0.305^{+0.007}_{-0.003}$ \\
\object[Bol 159]{B159}  & WFC/F475   & $26$       & K66  & $134.30$  & $-61.64\pm20.05$  & $8.90^{+0.30}_{-0.40}$  & $2.09^{+0.08}_{-0.11}$  & $15.34^{+0.06}_{-0.04}$  & $-1.010^{+0.028}_{-0.020}$  & $-0.430^{+0.028}_{-0.020}$ \\
          ~~        & ~~      & & W  & $174.01$  & $-93.00\pm22.10$  & $9.50^{+0.40}_{-0.40}$  & $3.32^{+0.01}_{-0.00}$  & $15.37^{+0.04}_{-0.04}$  & $-0.993^{+0.019}_{-0.020}$  & $-0.414^{+0.019}_{-0.020}$ \\
          ~~        & ~~     & & PL  & $351.18$  & $-36.68\pm2.33$       & ---  & $3.10^{+0.03}_{-0.00}$  & $15.56^{+0.01}_{-0.00}$  & $-0.905^{+0.010}_{-0.000}$  & $-0.326^{+0.010}_{-0.000}$ \\
          ~~        & ~~    & & K62  & $283.89$  & $-43.44\pm5.17$       & ---  & $3.18^{+0.02}_{-0.74}$  & $15.51^{+0.02}_{-0.00}$  & $-0.943^{+0.007}_{-0.001}$  & $-0.363^{+0.007}_{-0.001}$ \\
\object[Bol 159]{B159}  & WFC/F814  & $26$       & K66   & $29.94$  & $22.68\pm13.52$  & $8.60^{+0.20}_{-0.20}$  & $2.01^{+0.06}_{-0.06}$  & $13.73^{+0.03}_{-0.03}$  & $-1.027^{+0.015}_{-0.014}$  & $-0.448^{+0.015}_{-0.014}$ \\
          ~~        & ~~      & & W   & $44.15$  & $-16.39\pm21.23$  & $9.00^{+0.30}_{-0.30}$  & $3.33^{+0.00}_{-0.00}$  & $13.76^{+0.03}_{-0.03}$  & $-1.007^{+0.017}_{-0.016}$  & $-0.428^{+0.017}_{-0.016}$ \\
          ~~        & ~~     & & PL  & $108.08$  & $34.12\pm2.01$       & ---  & $3.10^{+0.02}_{-0.00}$  & $13.88^{+0.01}_{-0.00}$  & $-0.951^{+0.006}_{-0.000}$  & $-0.371^{+0.006}_{-0.000}$ \\
          ~~        & ~~    & & K62   & $80.51$  & $25.08\pm5.02$       & ---  & $3.18^{+0.02}_{-0.56}$  & $13.84^{+0.00}_{-0.00}$  & $-0.986^{+0.002}_{-0.001}$  & $-0.406^{+0.002}_{-0.001}$ \\
\object[SKHV 215]{B161-G215}  & WFC/F475   & $26$       & K66  & $178.91$  & $245.69\pm2.31$  & $14.70^{+0.70}_{-0.50}$  & $3.29^{+0.15}_{-0.11}$  & $10.21^{+0.23}_{-0.32}$  & $-2.623^{+0.097}_{-0.136}$  & $-2.043^{+0.097}_{-0.136}$ \\
          ~~        & ~~      & & W  & $411.91$  & $211.42\pm16.91$  & $7.70^{+0.30}_{-0.30}$  & $2.95^{+0.20}_{-0.25}$  & $14.65^{+0.07}_{-0.06}$  & $-0.910^{+0.031}_{-0.030}$  & $-0.330^{+0.031}_{-0.030}$ \\
          ~~        & ~~     & & PL  & $412.17$  & $203.67\pm6.14$       & ---  & $3.10^{+0.06}_{-0.00}$  & $14.64^{+0.04}_{-0.00}$  & $-0.927^{+0.027}_{-0.000}$  & $-0.347^{+0.027}_{-0.000}$ \\
          ~~        & ~~    & & K62  & $349.52$  & $227.37\pm20.87$       & ---  & $1.92^{+0.32}_{-0.20}$  & $14.64^{+0.05}_{-0.04}$  & $-0.938^{+0.023}_{-0.020}$  & $-0.359^{+0.023}_{-0.020}$ \\
\object[SKHV 215]{B161-G215}  & WFC/F814  & $26$       & K66   & $85.07$  & $314.08\pm7.86$  & $7.40^{+0.20}_{-0.20}$  & $1.65^{+0.06}_{-0.06}$  & $13.52^{+0.05}_{-0.05}$  & $-0.885^{+0.023}_{-0.022}$  & $-0.305^{+0.023}_{-0.022}$ \\
          ~~        & ~~      & & W  & $144.30$  & $293.95\pm16.11$  & $7.20^{+0.30}_{-0.30}$  & $2.54^{+0.25}_{-0.23}$  & $13.60^{+0.05}_{-0.05}$  & $-0.832^{+0.026}_{-0.025}$  & $-0.253^{+0.026}_{-0.025}$ \\
          ~~        & ~~     & & PL  & $156.42$  & $280.75\pm16.39$       & ---  & $3.25^{+0.10}_{-0.13}$  & $13.60^{+0.04}_{-0.08}$  & $-0.832^{+0.034}_{-0.053}$  & $-0.253^{+0.034}_{-0.053}$ \\
          ~~        & ~~    & & K62  & $109.03$  & $312.03\pm12.56$       & ---  & $1.64^{+0.12}_{-0.12}$  & $13.58^{+0.04}_{-0.04}$  & $-0.873^{+0.021}_{-0.017}$  & $-0.294^{+0.021}_{-0.017}$ \\
\object[SKHV 217]{B163-G217}  & WFPC/F555   & $40$       & K66  & $1231.00$  & $409.73\pm0.60$  & $8.80^{+0.30}_{-0.30}$  & $2.07^{+0.08}_{-0.08}$  & $12.30^{+0.27}_{-0.25}$  & $-1.597^{+0.091}_{-0.089}$  & $-1.018^{+0.091}_{-0.089}$ \\
          ~~        & ~~      & & W  & $190.45$  & $394.23\pm1.37$  & $8.70^{+0.20}_{-0.20}$  & $3.33^{+0.00}_{-0.02}$  & $10.72^{+0.27}_{-0.22}$  & $-1.914^{+0.087}_{-0.076}$  & $-1.335^{+0.087}_{-0.076}$ \\
          ~~        & ~~     & & PL  & $448.92$  & $389.51\pm0.88$       & ---  & $3.10^{+0.01}_{-0.00}$  & $11.74^{+0.02}_{-0.00}$  & $-1.593^{+0.008}_{-0.000}$  & $-1.013^{+0.008}_{-0.000}$ \\
          ~~        & ~~    & & K62  & $292.24$  & $386.31\pm1.52$       & ---  & $3.20^{+0.00}_{-0.10}$  & $11.47^{+0.00}_{-0.00}$  & $-1.697^{+0.001}_{-0.000}$  & $-1.118^{+0.001}_{-0.000}$ \\
\object[SKHV 217]{B163-G217}  & WFPC/F814  & $40$       & K66  & $366.09$  & $438.16\pm0.49$  & $8.60^{+0.30}_{-0.30}$  & $2.01^{+0.08}_{-0.09}$  & $11.91^{+0.31}_{-0.28}$  & $-1.417^{+0.097}_{-0.095}$  & $-0.838^{+0.097}_{-0.095}$ \\
          ~~        & ~~      & & W   & $55.82$  & $417.22\pm1.23$  & $8.70^{+0.10}_{-0.20}$  & $3.33^{+0.00}_{-0.02}$  & $10.10^{+0.31}_{-0.13}$  & $-1.803^{+0.095}_{-0.043}$  & $-1.224^{+0.095}_{-0.043}$ \\
          ~~        & ~~     & & PL  & $148.27$  & $412.90\pm1.07$       & ---  & $3.10^{+0.01}_{-0.00}$  & $11.16^{+0.03}_{-0.00}$  & $-1.474^{+0.010}_{-0.000}$  & $-0.895^{+0.010}_{-0.000}$ \\
          ~~        & ~~    & & K62   & $89.12$  & $408.35\pm0.82$       & ---  & $3.20^{+0.00}_{-0.10}$  & $10.83^{+0.11}_{-0.00}$  & $-1.589^{+0.023}_{-0.000}$  & $-1.010^{+0.023}_{-0.000}$ \\
\object[Bol 164]{B164-V253}  & WFC/F475   & $26$       & K66   & $40.59$  & $1073.86\pm8.29$  & $8.30^{+0.20}_{-0.10}$  & $1.92^{+0.06}_{-0.03}$  & $14.20^{+0.03}_{-0.05}$  & $-1.341^{+0.011}_{-0.021}$  & $-0.762^{+0.011}_{-0.021}$ \\
          ~~        & ~~      & & W   & $89.54$  & $1064.04\pm3.63$  & $14.70^{+0.50}_{-0.60}$  & $4.10^{+0.13}_{-0.15}$  & $9.82^{+0.23}_{-0.20}$  & $-2.950^{+0.103}_{-0.086}$  & $-2.371^{+0.103}_{-0.086}$ \\
          ~~        & ~~     & & PL  & $243.22$  & $1072.62\pm1.79$       & ---  & $3.10^{+0.02}_{-0.00}$  & $14.40^{+0.02}_{-0.00}$  & $-1.259^{+0.009}_{-0.000}$  & $-0.680^{+0.009}_{-0.000}$ \\
          ~~        & ~~    & & K62  & $166.10$  & $1062.44\pm5.38$       & ---  & $3.20^{+0.00}_{-0.52}$  & $14.33^{+0.00}_{-0.00}$  & $-1.301^{+0.002}_{-0.001}$  & $-0.722^{+0.002}_{-0.001}$ \\
\object[Bol 164]{B164-V253}  & WFC/F814  & $26$       & K66   & $23.43$  & $1796.27\pm13.16$  & $8.40^{+0.20}_{-0.20}$  & $1.95^{+0.06}_{-0.06}$  & $12.60^{+0.06}_{-0.07}$  & $-1.323^{+0.024}_{-0.027}$  & $-0.744^{+0.024}_{-0.027}$ \\
          ~~        & ~~      & & W   & $37.95$  & $1751.45\pm23.17$  & $8.50^{+0.20}_{-0.30}$  & $3.31^{+0.02}_{-0.07}$  & $12.70^{+0.06}_{-0.05}$  & $-1.271^{+0.028}_{-0.019}$  & $-0.691^{+0.028}_{-0.019}$ \\
          ~~        & ~~     & & PL   & $69.21$  & $1789.90\pm2.69$       & ---  & $3.10^{+0.02}_{-0.00}$  & $12.85^{+0.01}_{-0.00}$  & $-1.220^{+0.008}_{-0.000}$  & $-0.641^{+0.008}_{-0.000}$ \\
          ~~        & ~~    & & K62   & $51.69$  & $1776.92\pm8.59$       & ---  & $3.14^{+0.06}_{-0.52}$  & $12.78^{+0.01}_{-0.00}$  & $-1.265^{+0.004}_{-0.000}$  & $-0.685^{+0.004}_{-0.000}$ \\
\object[SKHV 233]{B182-G233}  & WFC/F555   & $28$       & K66  & $170.59$  & $45.49\pm6.65$  & $8.70^{+0.10}_{-0.10}$  & $2.04^{+0.03}_{-0.03}$  & $12.92^{+0.04}_{-0.04}$  & $-1.119^{+0.014}_{-0.014}$  & $-0.540^{+0.014}_{-0.014}$ \\
          ~~        & ~~      & & W  & $321.10$  & $23.98\pm5.82$  & $15.40^{+0.50}_{-0.50}$  & $4.28^{+0.13}_{-0.13}$  & $8.61^{+0.22}_{-0.23}$  & $-2.734^{+0.092}_{-0.095}$  & $-2.154^{+0.092}_{-0.095}$ \\
          ~~        & ~~     & & PL  & $959.51$  & $34.56\pm3.18$       & ---  & $3.10^{+0.02}_{-0.00}$  & $13.17^{+0.01}_{-0.00}$  & $-1.014^{+0.007}_{-0.000}$  & $-0.435^{+0.007}_{-0.000}$ \\
          ~~        & ~~    & & K62  & $649.07$  & $20.56\pm5.79$       & ---  & $3.20^{+0.00}_{-0.38}$  & $13.11^{+0.01}_{-0.01}$  & $-1.055^{+0.003}_{-0.002}$  & $-0.476^{+0.003}_{-0.002}$ \\
\object[SKHV 233]{B182-G233}  & WFC/F814  & $28$       & K66  & $169.93$  & $42.09\pm9.48$  & $7.90^{+0.20}_{-0.10}$  & $1.80^{+0.06}_{-0.03}$  & $12.84^{+0.01}_{-0.02}$  & $-0.845^{+0.009}_{-0.018}$  & $-0.266^{+0.009}_{-0.018}$ \\
          ~~        & ~~      & & W   & $95.37$  & $4.16\pm13.49$  & $7.80^{+0.20}_{-0.10}$  & $3.03^{+0.13}_{-0.08}$  & $12.87^{+0.01}_{-0.01}$  & $-0.814^{+0.007}_{-0.014}$  & $-0.235^{+0.007}_{-0.014}$ \\
          ~~        & ~~     & & PL  & $106.83$  & $-3.10\pm3.12$       & ---  & $3.10^{+0.02}_{-0.00}$  & $12.86^{+0.01}_{-0.00}$  & $-0.826^{+0.007}_{-0.000}$  & $-0.247^{+0.007}_{-0.000}$ \\
          ~~        & ~~    & & K62   & $98.03$  & $7.02\pm14.83$       & ---  & $2.28^{+0.26}_{-0.14}$  & $12.85^{+0.01}_{-0.01}$  & $-0.850^{+0.005}_{-0.006}$  & $-0.271^{+0.005}_{-0.006}$ \\
\object[SKHV 235]{B185-G235}  & WFC/F435   & $26$       & K66   & $80.06$  & $-39.52\pm5.04$  & $7.90^{+0.10}_{-0.10}$  & $1.80^{+0.03}_{-0.03}$  & $14.27^{+0.04}_{-0.04}$  & $-0.992^{+0.015}_{-0.015}$  & $-0.413^{+0.015}_{-0.015}$ \\
          ~~        & ~~      & & W  & $365.13$  & $-78.46\pm16.73$  & $7.80^{+0.20}_{-0.20}$  & $3.03^{+0.13}_{-0.16}$  & $14.36^{+0.04}_{-0.05}$  & $-0.945^{+0.019}_{-0.020}$  & $-0.365^{+0.019}_{-0.020}$ \\
          ~~        & ~~     & & PL  & $411.52$  & $-84.47\pm3.10$       & ---  & $3.10^{+0.02}_{-0.00}$  & $14.36^{+0.01}_{-0.00}$  & $-0.954^{+0.009}_{-0.000}$  & $-0.375^{+0.009}_{-0.000}$ \\
          ~~        & ~~    & & K62  & $318.60$  & $-66.09\pm19.08$       & ---  & $2.14^{+0.26}_{-0.16}$  & $14.33^{+0.03}_{-0.02}$  & $-0.980^{+0.012}_{-0.010}$  & $-0.400^{+0.012}_{-0.010}$ \\
\object[2MASX J00434551+4136578]{B193-G244}  & WFC/F475   & $26$       & K66  & $245.51$  & $164.21\pm7.20$  & $8.60^{+0.10}_{-0.20}$  & $2.01^{+0.03}_{-0.06}$  & $13.45^{+0.07}_{-0.04}$  & $-1.183^{+0.030}_{-0.017}$  & $-0.604^{+0.030}_{-0.017}$ \\
          ~~        & ~~      & & W  & $175.76$  & $115.55\pm13.29$  & $8.40^{+0.20}_{-0.10}$  & $3.30^{+0.02}_{-0.02}$  & $13.57^{+0.02}_{-0.04}$  & $-1.114^{+0.010}_{-0.021}$  & $-0.535^{+0.010}_{-0.021}$ \\
          ~~        & ~~     & & PL  & $832.24$  & $141.00\pm1.46$       & ---  & $3.10^{+0.01}_{-0.00}$  & $13.69^{+0.00}_{-0.00}$  & $-1.069^{+0.004}_{-0.000}$  & $-0.489^{+0.004}_{-0.000}$ \\
          ~~        & ~~    & & K62  & $426.87$  & $129.58\pm2.83$       & ---  & $3.18^{+0.02}_{-0.24}$  & $13.62^{+0.01}_{-0.00}$  & $-1.115^{+0.003}_{-0.000}$  & $-0.536^{+0.003}_{-0.000}$ \\
\object[SKHV 254]{B204-G254}  & WFC/F475   & $26$       & K66  & $289.58$  & $187.97\pm1.89$  & $15.50^{+0.50}_{-0.60}$  & $3.47^{+0.11}_{-0.13}$  & $9.14^{+0.29}_{-0.25}$  & $-2.847^{+0.119}_{-0.102}$  & $-2.267^{+0.119}_{-0.102}$ \\
          ~~        & ~~      & & W  & $688.16$  & $154.88\pm4.71$  & $14.40^{+0.60}_{-0.60}$  & $4.02^{+0.15}_{-0.15}$  & $9.40^{+0.23}_{-0.25}$  & $-2.677^{+0.105}_{-0.108}$  & $-2.098^{+0.105}_{-0.108}$ \\
          ~~        & ~~     & & PL  & $1048.10$  & $135.24\pm4.70$       & ---  & $3.10^{+0.03}_{-0.00}$  & $13.99^{+0.01}_{-0.00}$  & $-1.004^{+0.010}_{-0.000}$  & $-0.424^{+0.010}_{-0.000}$ \\
          ~~        & ~~    & & K62  & $881.25$  & $149.16\pm30.32$       & ---  & $2.22^{+0.84}_{-0.28}$  & $13.95^{+0.07}_{-0.07}$  & $-1.034^{+0.026}_{-0.025}$  & $-0.455^{+0.026}_{-0.025}$ \\
\object[SKHV 254]{B204-G254}  & WFC/F814  & $26$       & K66   & $67.27$  & $244.10\pm7.45$  & $7.50^{+0.20}_{-0.10}$  & $1.68^{+0.06}_{-0.03}$  & $12.75^{+0.02}_{-0.05}$  & $-0.936^{+0.013}_{-0.025}$  & $-0.357^{+0.013}_{-0.025}$ \\
          ~~        & ~~      & & W  & $180.25$  & $203.76\pm16.97$  & $7.40^{+0.20}_{-0.20}$  & $2.70^{+0.17}_{-0.17}$  & $12.79^{+0.03}_{-0.03}$  & $-0.898^{+0.018}_{-0.018}$  & $-0.319^{+0.018}_{-0.018}$ \\
          ~~        & ~~     & & PL  & $211.46$  & $185.85\pm16.09$       & ---  & $3.20^{+0.10}_{-0.05}$  & $12.79^{+0.05}_{-0.03}$  & $-0.903^{+0.038}_{-0.021}$  & $-0.323^{+0.038}_{-0.021}$ \\
          ~~        & ~~    & & K62  & $114.89$  & $232.06\pm10.53$       & ---  & $1.76^{+0.06}_{-0.08}$  & $12.77^{+0.02}_{-0.02}$  & $-0.935^{+0.012}_{-0.009}$  & $-0.355^{+0.012}_{-0.009}$ \\
\object[SKHV 264]{B213-G264}   & WFC/F435   & $26$       & K66   & $37.49$  & $6.48\pm1.65$  & $8.90^{+0.10}_{-0.20}$  & $2.09^{+0.03}_{-0.05}$  & $14.18^{+0.15}_{-0.09}$  & $-1.354^{+0.049}_{-0.029}$  & $-0.775^{+0.049}_{-0.029}$ \\
          ~~        & ~~      & & W   & $29.98$  & $0.33\pm0.65$  & $16.00^{+0.00}_{-0.20}$  & $4.44^{+0.00}_{-0.05}$  & $9.67^{+0.10}_{-0.00}$  & $-3.063^{+0.039}_{-0.000}$  & $-2.484^{+0.039}_{-0.000}$ \\
          ~~        & ~~     & & PL  & $514.94$  & $-1.95\pm1.01$       & ---  & $3.10^{+0.02}_{-0.00}$  & $14.81^{+0.04}_{-0.00}$  & $-1.129^{+0.016}_{-0.000}$  & $-0.550^{+0.016}_{-0.000}$ \\
          ~~        & ~~    & & K62  & $368.45$  & $-7.52\pm3.17$       & ---  & $3.18^{+0.02}_{-0.46}$  & $14.70^{+0.04}_{-0.01}$  & $-1.183^{+0.011}_{-0.001}$  & $-0.604^{+0.011}_{-0.001}$ \\
\object[SKHV 011]{B293-G011}  & WFPC/F555   & $29$       & K66   & $39.41$  & $7.67\pm7.46$  & $6.60^{+0.10}_{-0.10}$  & $1.41^{+0.03}_{-0.03}$  & $13.52^{+0.01}_{-0.01}$  & $-0.556^{+0.010}_{-0.010}$  & $0.023^{+0.010}_{-0.010}$ \\
          ~~        & ~~      & & W   & $45.85$  & $-27.12\pm9.33$  & $6.40^{+0.10}_{-0.10}$  & $2.00^{+0.06}_{-0.05}$  & $13.54^{+0.01}_{-0.01}$  & $-0.519^{+0.008}_{-0.008}$  & $0.060^{+0.008}_{-0.008}$ \\
          ~~        & ~~     & & PL   & $83.68$  & $-56.29\pm16.20$       & ---  & $3.60^{+0.10}_{-0.05}$  & $13.55^{+0.02}_{-0.01}$  & $-0.479^{+0.025}_{-0.013}$  & $0.100^{+0.025}_{-0.013}$ \\
          ~~        & ~~    & & K62   & $34.97$  & $5.86\pm8.13$       & ---  & $1.36^{+0.04}_{-0.02}$  & $13.51^{+0.01}_{-0.01}$  & $-0.570^{+0.005}_{-0.010}$  & $0.009^{+0.005}_{-0.010}$ \\
\object[SKHV 011]{B293-G011}  & WFPC/F606  & $29$       & K66  & $101.34$  & $4.74\pm2.87$  & $7.00^{+0.10}_{-0.20}$  & $1.53^{+0.03}_{-0.06}$  & $14.45^{+0.04}_{-0.02}$  & $-0.473^{+0.028}_{-0.014}$  & $0.106^{+0.028}_{-0.014}$ \\
          ~~        & ~~      & & W   & $35.01$  & $-6.38\pm4.23$  & $6.50^{+0.10}_{-0.20}$  & $2.05^{+0.06}_{-0.11}$  & $14.51^{+0.02}_{-0.01}$  & $-0.399^{+0.020}_{-0.010}$  & $0.180^{+0.020}_{-0.010}$ \\
          ~~        & ~~     & & PL   & $39.38$  & $-13.70\pm2.63$       & ---  & $3.65^{+0.05}_{-0.05}$  & $14.53^{+0.01}_{-0.01}$  & $-0.337^{+0.014}_{-0.015}$  & $0.243^{+0.014}_{-0.015}$ \\
          ~~        & ~~    & & K62   & $74.37$  & $3.60\pm3.67$       & ---  & $1.48^{+0.06}_{-0.04}$  & $14.45^{+0.02}_{-0.03}$  & $-0.474^{+0.012}_{-0.018}$  & $0.105^{+0.012}_{-0.018}$ \\
\object[SKHV 011]{B293-G011}  & WFPC/F814  & $29$       & K66   & $98.64$  & $12.62\pm4.14$  & $6.90^{+0.20}_{-0.10}$  & $1.50^{+0.06}_{-0.03}$  & $13.84^{+0.02}_{-0.03}$  & $-0.444^{+0.013}_{-0.027}$  & $0.135^{+0.013}_{-0.027}$ \\
          ~~        & ~~      & & W   & $52.74$  & $-1.96\pm5.49$  & $6.50^{+0.10}_{-0.20}$  & $2.05^{+0.06}_{-0.11}$  & $13.88^{+0.02}_{-0.01}$  & $-0.384^{+0.019}_{-0.010}$  & $0.195^{+0.019}_{-0.010}$ \\
          ~~        & ~~     & & PL   & $53.80$  & $-10.22\pm6.76$       & ---  & $3.65^{+0.10}_{-0.10}$  & $13.90^{+0.02}_{-0.02}$  & $-0.322^{+0.026}_{-0.030}$  & $0.257^{+0.026}_{-0.030}$ \\
          ~~        & ~~    & & K62   & $80.79$  & $9.51\pm4.82$       & ---  & $1.48^{+0.04}_{-0.06}$  & $13.83^{+0.02}_{-0.01}$  & $-0.459^{+0.018}_{-0.011}$  & $0.121^{+0.018}_{-0.011}$ \\
\object[SKHV 307]{B375-G307}  & WFC/F475   & $26$       & K66   & $25.94$  & $155.47\pm1.67$  & $8.60^{+0.20}_{-0.10}$  & $2.01^{+0.06}_{-0.03}$  & $14.84^{+0.06}_{-0.12}$  & $-1.190^{+0.020}_{-0.042}$  & $-0.611^{+0.020}_{-0.042}$ \\
          ~~        & ~~      & & W   & $29.69$  & $151.23\pm1.21$  & $15.60^{+0.40}_{-0.50}$  & $4.33^{+0.10}_{-0.13}$  & $10.29^{+0.24}_{-0.19}$  & $-2.899^{+0.097}_{-0.077}$  & $-2.319^{+0.097}_{-0.077}$ \\
          ~~        & ~~     & & PL  & $187.86$  & $148.78\pm0.81$       & ---  & $3.10^{+0.02}_{-0.00}$  & $15.23^{+0.01}_{-0.00}$  & $-1.040^{+0.008}_{-0.000}$  & $-0.461^{+0.008}_{-0.000}$ \\
          ~~        & ~~    & & K62  & $128.81$  & $145.30\pm1.96$       & ---  & $3.20^{+0.00}_{-0.48}$  & $15.12^{+0.03}_{-0.00}$  & $-1.094^{+0.009}_{-0.000}$  & $-0.515^{+0.009}_{-0.000}$ \\
\object[SKHV 307]{B375-G307}  & WFC/F814  & $26$       & K66   & $22.19$  & $138.54\pm1.56$  & $8.60^{+0.20}_{-0.10}$  & $2.01^{+0.06}_{-0.03}$  & $13.80^{+0.04}_{-0.09}$  & $-1.166^{+0.016}_{-0.037}$  & $-0.587^{+0.016}_{-0.037}$ \\
          ~~        & ~~      & & W   & $23.87$  & $122.78\pm3.38$  & $8.60^{+0.10}_{-0.20}$  & $3.32^{+0.01}_{-0.02}$  & $13.87^{+0.05}_{-0.03}$  & $-1.120^{+0.023}_{-0.012}$  & $-0.541^{+0.023}_{-0.012}$ \\
          ~~        & ~~     & & PL   & $88.54$  & $132.29\pm0.74$       & ---  & $3.10^{+0.02}_{-0.00}$  & $14.05^{+0.01}_{-0.00}$  & $-1.048^{+0.008}_{-0.000}$  & $-0.468^{+0.008}_{-0.000}$ \\
          ~~        & ~~    & & K62   & $55.77$  & $129.26\pm1.12$       & ---  & $3.18^{+0.02}_{-0.30}$  & $13.98^{+0.00}_{-0.00}$  & $-1.094^{+0.002}_{-0.001}$  & $-0.514^{+0.002}_{-0.001}$ \\
\object[CXO J004246.0+411736]{BH16}  & WFC/F435   & $48$       & K66   & $47.35$  & $770.54\pm0.55$  & $9.20^{+0.30}_{-0.20}$  & $2.17^{+0.07}_{-0.05}$  & $14.39^{+0.17}_{-0.29}$  & $-1.682^{+0.058}_{-0.099}$  & $-1.103^{+0.058}_{-0.099}$ \\
          ~~        & ~~      & & W   & $31.80$  & $757.13\pm2.30$  & $9.30^{+0.20}_{-0.20}$  & $3.32^{+0.00}_{-0.00}$  & $14.57^{+0.13}_{-0.15}$  & $-1.608^{+0.047}_{-0.054}$  & $-1.028^{+0.047}_{-0.054}$ \\
          ~~        & ~~     & & PL  & $104.59$  & $773.77\pm0.59$       & ---  & $3.10^{+0.02}_{-0.00}$  & $15.22^{+0.02}_{-0.00}$  & $-1.375^{+0.010}_{-0.000}$  & $-0.796^{+0.010}_{-0.000}$ \\
          ~~        & ~~    & & K62   & $83.00$  & $771.05\pm0.99$       & ---  & $3.20^{+0.00}_{-0.30}$  & $15.11^{+0.00}_{-0.00}$  & $-1.430^{+0.001}_{-0.000}$  & $-0.850^{+0.001}_{-0.000}$ \\
\object[NBol 021]{NB21}  & WFC/F435   & $26$       & K66   & $27.75$  & $1562.70\pm4.95$  & $1.00^{+1.60}_{-0.00}$  & $0.30^{+0.31}_{-0.00}$  & $18.76^{+0.00}_{-0.01}$  & $0.042^{+0.000}_{-0.204}$  & $0.621^{+0.000}_{-0.204}$ \\
          ~~        & ~~      & & W   & $28.76$  & $1556.46\pm5.06$  & $1.00^{+1.90}_{-0.00}$  & $0.56^{+0.39}_{-0.00}$  & $18.75^{+0.00}_{-0.01}$  & $0.070^{+0.000}_{-0.226}$  & $0.649^{+0.000}_{-0.226}$ \\
          ~~        & ~~     & & PL   & $27.19$  & $1562.38\pm8.07$       & ---  & $13.00^{+0.00}_{-6.70}$  & $18.77^{+0.00}_{-0.02}$  & $0.159^{+0.000}_{-0.212}$  & $0.738^{+0.000}_{-0.212}$ \\
          ~~        & ~~    & & K62   & $29.21$  & $1560.01\pm4.95$       & ---  & $0.50^{+0.14}_{-0.00}$  & $18.75^{+0.00}_{-0.01}$  & $-0.199^{+0.000}_{-0.047}$  & $0.380^{+0.000}_{-0.047}$ \\
\object[NBol 021]{NB21}  & WFC/F475   & $25$       & K66   & $13.26$  & $2581.38\pm13.73$  & $9.40^{+0.30}_{-0.30}$  & $2.22^{+0.07}_{-0.07}$  & $15.76^{+0.02}_{-0.02}$  & $-1.206^{+0.010}_{-0.009}$  & $-0.626^{+0.010}_{-0.009}$ \\
          ~~        & ~~      & & W   & $16.47$  & $2557.48\pm16.22$  & $10.30^{+0.40}_{-0.40}$  & $3.34^{+0.03}_{-0.01}$  & $15.77^{+0.02}_{-0.02}$  & $-1.200^{+0.012}_{-0.010}$  & $-0.621^{+0.012}_{-0.010}$ \\
          ~~        & ~~     & & PL   & $78.07$  & $2620.20\pm2.13$       & ---  & $3.10^{+0.03}_{-0.00}$  & $15.92^{+0.01}_{-0.00}$  & $-1.124^{+0.007}_{-0.000}$  & $-0.545^{+0.007}_{-0.000}$ \\
          ~~        & ~~    & & K62   & $61.43$  & $2614.76\pm2.99$       & ---  & $3.18^{+0.02}_{-0.66}$  & $15.88^{+0.00}_{-0.00}$  & $-1.156^{+0.003}_{-0.000}$  & $-0.577^{+0.003}_{-0.000}$ \\
\object[NBol 021]{NB21}  & WFC/F814  & $25$       & K66   & $15.58$  & $5218.87\pm74.17$  & $10.80^{+1.30}_{-1.00}$  & $2.51^{+0.25}_{-0.20}$  & $14.40^{+0.04}_{-0.02}$  & $-1.195^{+0.021}_{-0.010}$  & $-0.616^{+0.021}_{-0.010}$ \\
          ~~        & ~~      & & W   & $15.93$  & $5196.36\pm76.45$  & $11.80^{+2.00}_{-1.00}$  & $3.49^{+0.39}_{-0.12}$  & $14.41^{+0.04}_{-0.03}$  & $-1.191^{+0.019}_{-0.016}$  & $-0.612^{+0.019}_{-0.016}$ \\
          ~~        & ~~     & & PL   & $41.72$  & $5338.91\pm4.98$       & ---  & $3.10^{+0.05}_{-0.00}$  & $14.60^{+0.01}_{-0.00}$  & $-1.085^{+0.012}_{-0.000}$  & $-0.506^{+0.012}_{-0.000}$ \\
          ~~        & ~~    & & K62   & $36.56$  & $5330.30\pm8.51$       & ---  & $3.20^{+0.00}_{-0.94}$  & $14.58^{+0.00}_{-0.00}$  & $-1.114^{+0.004}_{-0.001}$  & $-0.535^{+0.004}_{-0.001}$ 
\enddata
\end{deluxetable}
\end{document}